# On the H-atom abstractions from $C_1$-$C_4$ alcohols, aldehydes, and ethers by $NO_2$: *ab initio* and comprehensive kinetic modeling.

Hongqing Wu[a,^], Ruoyue Tang[a,^], Yuxin Dong[a], Xinrui Ren[a], Mingrui Wang[a], Ting Zhang[a], Hongjie Zhang[a], Guangyuan Feng[a], Song Cheng[a,b]*

[a] *Department of Mechanical Engineering, The Hong Kong Polytechnic University, Kowloon, Hong Kong SAR, China*
[b] *Research Institute for Smart Energy, The Hong Kong Polytechnic University, Kowloon, Hong Kong SAR, China*

^ Both authors contribute equally to this paper.

* Corresponding authors: Song Cheng

Phone: +852 2766 6668

Email: songcheng@polyu.edu.hk

## Novelty and Significance Statement

This study provides an extensive investigation into H-atom abstraction by $NO_2$ from $C_1$-$C_4$ alcohols, aldehydes, and ethers, involving 9 hydrocarbons and 3 $HNO_2$ isomers across 45 reactions. By utilizing quantum chemistry computations, this study obtained all kinetic parameters, based on which branching ratios and rate rules for these reactions on oxygenated species are established. These rules enable reliable derivation of rate parameters for a broader spectrum of hydrocarbons. The chemical kinetic modeling and comprehensive analysis further demonstrate the significant impact of these reactions on $NO_X$/hydrocarbon interactions and model reactivity. The findings underscore the importance of accurately representing their kinetics in current models, a goal that can now be achieved with the data presented in this research.

## CRediT authorship contribution statement

**Hongqing Wu**: Writing – original draft; Conceptualization; Methodology; Software; Formal analysis; Investigation. **Ruoyue Tang**: Writing – original draft; Conceptualization; Methodology; Software; Formal analysis; Investigation. **Yuxin Dong**: Methodology; Software; Writing - Review & Editing. **Xinrui Ren**: Methodology; Software; Writing - Review & Editing. **Mingrui Wang**: Methodology; Software; Writing - Review & Editing. **Ting Zhang**: Methodology; Software; Writing - Review & Editing. **Hongjie Zhang**: Methodology; Software; Writing - Review & Editing. **Guangyuan Feng**: Methodology; Software; Writing - Review & Editing. **Song Cheng**: Writing – original draft; Conceptualization; Methodology; Software; Formal analysis; Investigation; Writing - Review & Editing; Supervision; Project administration; Funding acquisition.

## Abstract:

As crucial additives and intermediate, alcohols, ethers, and aldehydes play a significant role in the combustion process. However, the chemistry of $NO_X$/hydrocarbon interactions and the rate rules governing these interactions remain largely unexplored in this combustion system. To address this gap, this study provides a comprehensive investigation of H-atom abstraction by $NO_2$ from $C_1$-$C_4$ alcohols, aldehydes and ethers that leads to the formation of 3 $HNO_2$ isomers (i.e., *trans*-HONO, $HNO_2$, and *cis*-HONO), encompassing 9 hydrocarbons and 45 reactions. Utilizing the DLPNO-CCSD(T)/cc-pVDZ//M06–2X/6−311++g(d,p) method, the electronic structures, single point energies, C-H bond dissociation energies and 1-D hindered rotor potentials of the reactants, transition states, complexes and products in each reaction are computed. The potential energy surfaces and energy barriers for each reaction are determined based on these calculations. Subsequently, the rate coefficients for all studied reactions are derived using transition state theory, implemented with the Master Equation System Solver program, across a temperature range from 298.15 to 2000 K. A thorough analysis of branching ratios highlights the differences and similarities between species, $HNO_2$ isomers, and abstraction sites, leading to the establishment of consistent rate rules that can be used for rate estimation by analogy for wider range of oxygenated species. Adding these H-atom abstractions to the chemical kinetic model improves the model reactivity and advances the ignition, as indicated by the reduction in ignition delay time for species that initially lacked these reactions. Further sensitivity and flux analyses highlight the crucial role of H-atom abstraction by $NO_2$. The findings underscore the importance of accurately incorporating these kinetic parameters into newly developed chemical models for alcohols, aldehydes, and ethers. Additionally, the study highlights the need for future experimental efforts to investigate the effects of $NO_2$ on the combustion systems of these compounds.

Keywords: *H-atom abstraction reaction by $NO_2$; ab initio calculations; $NO_x$ interaction*

# 1. Introduction

Understanding the formation and consumption of nitrogen oxides ($NO_X$) in combustion process has become more urgent than ever. On one hand, alcohol and ether, as renewable and economical green fuels, are being used as additives or promising alternatives to petroleum-based fuels. Their application can improve combustion performance and further reduce the emission of $NO_X$ [1-4]. On the other hand, exhaust gas recirculation (EGR) technology is widely applied in internal combustion processes, enabling better and wider control of combustion phasing/heat release rates (HRRs) in homogeneous charge compression ignition engines [5], as well as knock in spark-ignition engines [6]. Although the $NO_X$ species are the minor (e.g., 10 - 250 ppm) in the mixture, they can significantly alter the original reaction pathways, leading to changes in system reactivity [7].

As such, in recent years, the interactions between $NO_X$ species and typical fuel components have been extensively investigated [8-13]. At the same time, research has revealed the complex mechanisms by which $NO_X$ either promotes or inhibits fuel reactivity. However, there are some limitations based on these studies: (a) Prior research has predominantly focused on NO due to its higher concentration, overlooking the rapid interconversion between $NO_2$ and NO [14]. (b) While existing studies have just considered alkanes and aromatics, they often neglect the roles of alcohols and ethers as additives or fuels, and aldehydes as the important intermediates. (c) Earlier works have concentrated on simple compounds (e.g., HCHO, $CH_3OH$, $CH_3OCH_3$), lacking comprehensive and systematic research on the impact of alcohols, aldehydes, and ethers. Dayma et al. [15] conducted experimental and modeling analysis to explore the kinetics of the interaction between $NO_2$ and methanol in a spherical fused silica jet-stirred reactor (JSR), discovering that direct interactions between $NO_2$ and fuel molecules or their primary derivatives significantly enhance reactivity. Similarly, Xiao et al. [16] investigated the high-pressure ignition characteristics of $NO_2/CH_4$ mixtures in a gas flow reactor, finding that $NO_2$ significantly reduced the ignition delay time of methanol. Very recently, Cheng et al. [7] characterized the interactions between $NO_X$ and ethylene/propylene/isobutylene in a rapid compression machine, revealing the strong influences of these interactions on the autoignition reactivity of gasoline fuels. Through limited, these studies have consistently revealed a major type of interaction reactions directly involving $NO_2$, namely $RH+NO_2 = \dot{R}+HNO_2/HONO$, that greatly promotes reactivity.

There have been a few experimental and theoretical studies to determine the rate coefficients of $RH+NO_2 = \dot{R}+HNO_2/HONO$. Experimental data are available for the reactions of $CH_3OH+NO_2$ at temperatures ranging from 639-713K [17]and 900-1100K [18], and for $HCHO+NO_2$ at 391-457K [19], 393-476K [20], and 1140-1650K [21]. However, these data are available over limited and narrow temperature ranges. In theoretical studies, Xiao et al. [16] calculated the rate coefficients for H-atom abstraction from $CH_3OH$ at the CBS-Q//B3LYP/6-311++G(2d,p) level of theory. Xu et al. [22] studied the reaction kinetics for all abstraction channels of HCHO + $NO_2$ using G2M//B3LYP/6-311+G(d,p) and CVT/SCT calculations. Gao et al. [23] employed CCSD(T)/CBS//B3LYP/6–311 + G(3DF, 3P) method to compute the rate coefficients of HCHO + $NO_2$. Wu et al. [1] calculated the rate coefficients for H-atom abstraction from $CH_3OH$ and HCHO by $NO_2$ at the CCSD(T)/aug-cc-pVQZ//M06–2X-D3/6-311++G(d,p) level of theory. Shang et al. [2] calculated the H-atom abstraction from several ethers by $NO_2$ at G04//B3LYP/6-31G(2df,p) and G4MP2// B3LYP/6-31G(2df,p) level of theory. Shi et al. [24] exploited kinetics study of the $CH_3OCH_3$ + $NO_2$ and $CH_3OC_2H_5$ + $NO_2$ reactions by CCSD(T)/CBS//M06-2X/ccpVTZ method. More recently, Wu et al. [25] determined the rate coefficients for H-atom abstractions by $NO_2$ from $C_2$-$C_5$ alkanes and alkenes at the DLPNO-CCSD(T)/cc-pVDZ//M06–2X/6−311++g(d,p) level of theory, while Guo et al. [26] investigated similar reactions for $C_3$-$C_7$ alkynes and dienes. Nevertheless, previous studies have largely focused on simple compounds, resulting in a lack of systematic analysis of alcohols, aldehydes, and ethers. Moreover, the absence of rate coefficients for these reactions introduces significant uncertainties in existing model predictions.

Therefore, this study aims to address these gaps by: (a) conducting a detailed theoretical investigation of H-atom abstractions from $C_1$-$C_4$ alcohols, aldehydes, and ethers by $NO_2$, leading to the formation of $HNO_2$, *trans*-HONO, and *cis*-HONO, while considering the formation of complexes where applicable; (b) revealing the branching ratios of the three pathways forming the three $HNO_2$ isomers for the selected species at different H-atom sites and molecule sizes; (c) determining rate rules for this type of reactions for alcohols, aldehydes, and ethers; and (d) systematically analyzing the effects of these reactions on model predictions.

## 2. Computational methods

### *2.1. Potential energy surfaces*

Electronic geometries, vibrational frequencies and zero-point energies for all species involved in the 24 reactions (including reactants, products, complexes, transition states (TS)) are calculated at the M06-2X method [27] coupled with the 6-311++G(d,p) basis set [28-30]. Conformer search at the same level of theory is conducted to ensure the optimized structures retain the lowest energy. Intrinsic reaction coordinate (IRC) calculations have been carried out at the same level of theory to ensure that the transition state connects the respective reactants with the respective product complex. 1-D hindered rotor treatment [31] is also obtained at the M06–2X/6–311++G(d, p) level of theory for the low frequency torsional modes between non-hydrogen atoms in all of the reactants, TSs, complexes and products, with a total of 18 scans (i.e., 20 degrees increment in the respective dihedral angle) for each rotor. Scale factors of 0.983 for harmonic frequencies and 0.9698 for ZPEs that were recommended by Zhao and Truhlar [27] are used herein. Single-point energies (SPEs) are further determined for all the species using the DLPNO-CCSD(T) functional [32,33] with the cc-pVDZ basis set. With the CCSD(T) method, attention must be addressed to T1 diagnostic [34] to measure the multi-reference effect. The T1 diagnostic values for all the species, as summarized in Table S1 in the Supplementary Material, are below 0.036, which indicates that the SPEs calculated from using single-reference calculation method are reliable in this study. All the calculations mentioned above are performed using ORCA 5.0.4 [35], and the optimized structures for all species, TSs and complexes are summarized in the Supplementary Material.

### *2.2. Rate constant calculations*

The Master Equation System Solver (MESS) program suite [36] is employed here to calculate the chemical rate coefficients for complex-forming reactions via solution of the one-dimensional master equation, based on the chemically significant eigenstate approach of Miller and Klippenstein [37] and the bimolecular species model of Georgievskii and Klippenstein [38]. The frequencies of lower-frequency modes are replaced by the hindered rotor potentials obtained from 1-D scans. Quantum mechanical tunneling corrections assuming the asymmetric Eckart potential (TST/Eck) [39] are applied to obtain the rate coefficient over the temperature range of 298–2000 K. All rate coefficients were fitted to the modified Arrhenius equation, which can be defined as $k = AT^n \exp(-E_a/RT)$.

### *2.3. Kinetic modeling*

The latest chemistry model developed by LLNL [40] is utilized in this study to investigate the impact of these calculated reactions on the model's performance in predicting the combustion characteristics of alcohols, aldehydes, and ethers. The development of this chemistry model has been documented in [40], and will not be detailed here. Kinetic modeling of fundamental combustion experiments, specifically autoignition experiments, was performed using LLNL's fast solver, Zero-RK [41]

# 3. Results and discussion

## *3.1. Species and reaction sites*

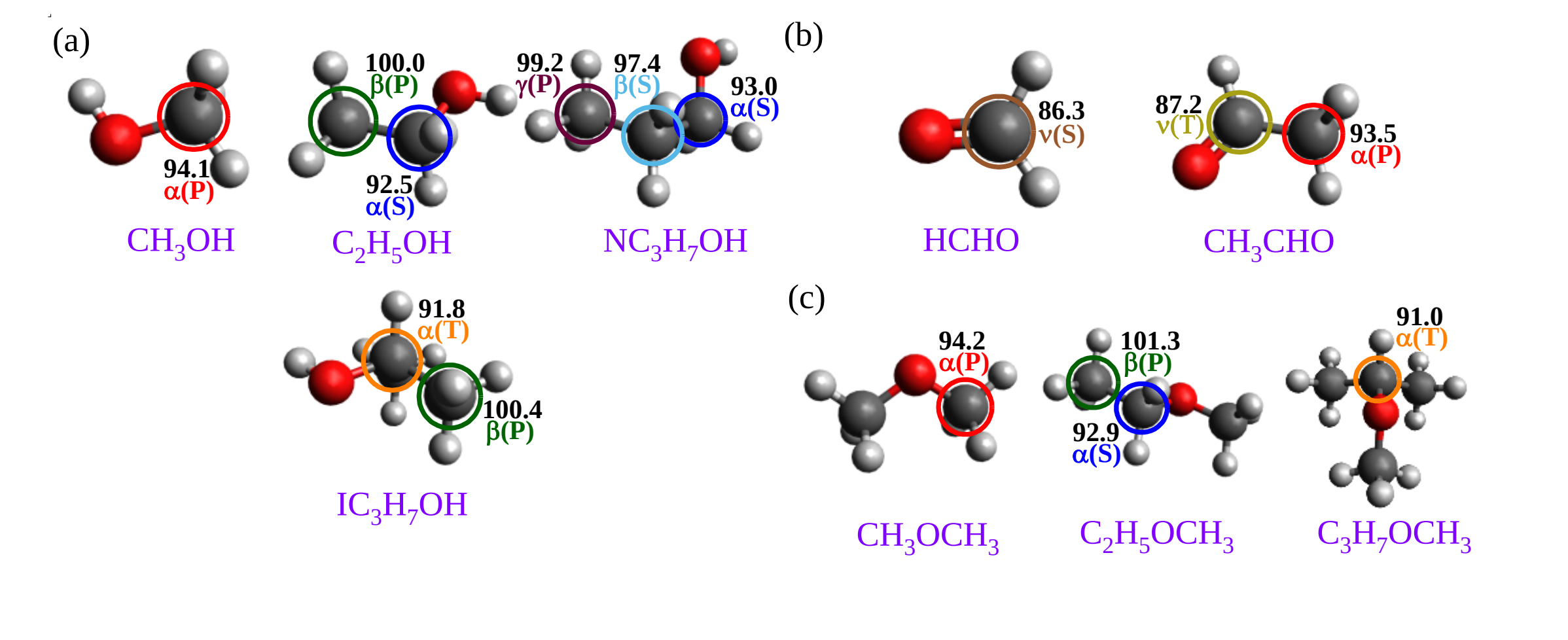

Figure 1. Different abstraction sites and bond dissociation energies (*kcal/mol*) of (a) alcohols, (b) aldehydes and (c)ethers.

This study involves the investigation of 45 reactions, in which the H atom is abstracted from 4 alcohols ($CH_3OH$, $C_2H_5OH$, $NC_3H_7OH$, and $IC_3H_7OH$), 2 aldehydes (HCHO and $CH_3CHO$), and 3 ethers ($CH_3OCH_3$, $C_2H_5OCH_3$, and $C_3H_7OCH_3$) by $NO_2$, leading to the formation of three $HNO_2$ isomers: *trans*-HONO, $HNO_2$, and *cis*-HONO. To better illustrate these reactions, the reaction sites for each species are marked in Fig. 1. The selection of reaction species and sites is to achieve a reasonable representation in carbon site type and carbon chain structure, so that the rate of this type of reaction for various alcohol, aldehyde and ether species can be determined using the results in this study by analogy. According to the type of C-atoms to which H-atoms bond, the C atom sites are classified into primary site (P), secondary site (S), and tertiary site (T). Additionally, reaction sites are designated as α, β, and γ according to the proximity to the functional group, while those located at C=O bonds are defined as ν.

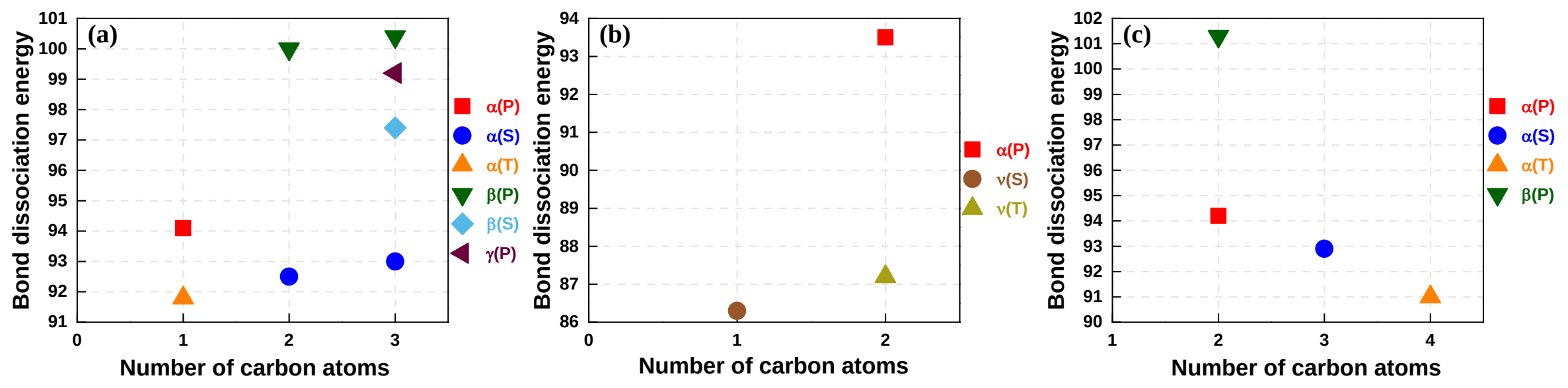

Figure 2. Bond dissociation energy of (a) alcohols, (b) aldehydes, and (c) ethers versus carbon number.

Furthermore, C-H bond dissociation energies (BDEs) for different reaction sites at 298 K are also calculated via

$$BDE_{298}(R-H) = \Delta H^0_{f,298}(\dot{R},g) + \Delta H^0_{f,298}(\dot{H},g) - \Delta H^0_{f,298}(RH,g)$$

The BDEs at different reaction sites of alcohols, aldehydes and ethers are illustrated in Fig.1. Then the BDEs at different reaction sites of alcohols, aldehydes and ethers are plotted as functions of carbon number, which are summarized in Fig. 2. As depicted in Figs. 2(a) and 2(c), the BDEs of alcohols follow the range from highest to the lowest is β(P) > γ(P) > β(S) > α(P) > α(S) > α(T), while for ethers, it is β(P) > α(P) > α(S) > α(T). These trends exhibit a notable similarity, reinforcing findings from the author's previous research [25]. For aldehydes (as shown in Fig. 2(b)), the BDEs at the ν(S) and ν(T) sites are nearly identical, with a difference of less than 0.9 kcal/mol. On the contrary, the BDE at the α(P) site is significantly higher than those at the ν site, with a discrepancy of approximately 6 kcal/mol. This pronounced difference indicates that the C=O functional group in aldehydes reduces the BDE at the ν stie.

### 3.2. *Potential energy surface*

Table 1. The relative energy for H-atom abstraction by $NO_2$ from different sites of alcohols, aldehydes and ethers to form the respective products and $HNO_2$ isomers (*trans*-HONO, $HNO_2$, *cis*-HONO). All values are in *kcal/mol*.

| No. | Reaction | Reactant | Transition state | Product complex | Product |
|---|---|---|---|---|---|
| | Alcohols + $NO_2$ | | | | |
| R1 | $CH_3OH$+$NO_2$ = $CH_2OH$+*trans*-HONO | 0 | 30.7 | 12.5 | 17.0 |
| R2 | $CH_3OH$ +$NO_2$ = $CH_2OH$+$HNO_2$ | 0 | 25.7 | 20.3 | 26.1 |
| R3 | $CH_3OH$ +$NO_2$ = $CH_2OH$+*cis*-HONO | 0 | 24.6 | 13.9 | 17.4 |
| R4 | $C_2H_5OH$+$NO_2$ = $C_2H_4OH$_α(S)+*trans*-HONO | 0 | 28.2 | 10.3 | 15.5 |
| R5 | $C_2H_5OH$ +$NO_2$ = $C_2H_4OH$_α(S)+$HNO_2$ | 0 | 22.6 | 18.0 | 24.5 |
| R6 | $C_2H_5OH$ +$NO_2$ = $C_2H_4OH$_α(S)+*cis*-HONO | 0 | 20.6 | 10.6 | 15.8 |
| R7 | $C_2H_5OH$ +$NO_2$ = $C_2H_4OH$_β(P)+*trans*-HONO | 0 | 34.3 | 15.7 | 23.0 |
| R8 | $C_2H_5OH$ +$NO_2$ = $C_2H_4OH$_β(P)+$HNO_2$ | 0 | 30.2 | 24.8 | 32.0 |
| R9 | $C_2H_5OH$ +$NO_2$ = $C_2H_4OH$_β(P)+*cis*-HONO | 0 | 29.3 | 16.6 | 23.3 |
| R10 | $NC_3H_7OH$+$NO_2$ = $NC_3H_6OH$_α(S)+*trans*-HONO | 0 | 29.6 | 11.1 | 16.0 |
| R11 | $NC_3H_7OH$+$NO_2$ = NC3H6OH_α(S)+$HNO_2$ | 0 | 22.6 | 18.2 | 25.0 |
| R12 | $NC_3H_7OH$+$NO_2$ = NC3H6OH_α(S)+*cis*-HONO | 0 | 22.2 | 11.6 | 16.3 |
| R13 | $NC_3H_7OH$+$NO_2$ = NC3H6OH_β(S)+*trans*-HONO | 0 | 32.3 | 16.8 | 20.3 |
| R14 | $NC_3H_7OH$+$NO_2$ = NC3H6OH_β(S)+$HNO_2$ | 0 | 26.5 | 21.1 | 29.4 |
| R15 | $NC_3H_7OH$+$NO_2$ = NC3H6OH_β(S)+*cis*-HONO | 0 | 25.6 | 13.1 | 20.6 |
| R16 | $NC_3H_7OH$+$NO_2$ = $NC_3H_6OH$_γ(P)+*trans*-HONO | 0 | 34.4 | 14.5 | 22.1 |
| R17 | $NC_3H_7OH$+$NO_2$ = $NC_3H_6OH$_γ(P)+$HNO_2$ | 0 | 28.8 | 22.3 | 31.1 |
| R18 | $NC_3H_7OH$+$NO_2$ = $NC_3H_6OH$_γ(P)+*cis*-HONO | 0 | 28.5 | 15.6 | 22.4 |
| R19 | $IC_3H_7OH$+$NO_2$ = $IC_3H_6OH$_α(T)+*trans*-HONO | 0 | 28.1 | 9.3 | 14.8 |
| R20 | $IC_3H_7OH$ +$NO_2$ = $IC_3H_6OH$_α(T)+$HNO_2$ | 0 | 20.2 | 16.6 | 23.8 |
| R21 | $IC_3H_7OH$ +$NO_2$ = $IC_3H_6OH$_α(T)+*cis*-HONO | 0 | 18.7 | 9.1 | 15.1 |
| R22 | $IC_3H_7OH$+$NO_2$ = $IC_3H_6OH$_β(P)+*trans*-HONO | 0 | 34.4 | 15.9 | 23.3 |
| R23 | $IC_3H_7OH$ +$NO_2$ = $IC_3H_6OH$_β(P)+$HNO_2$ | 0 | 30.5 | 24.8 | 32.4 |
| R24 | $IC_3H_7OH$ +$NO_2$ = $IC_3H_6OH$_β(P)+*cis*-HONO | 0 | 28.9 | 17.2 | 23.6 |
| | Aldehydes + $NO_2$ | | | | |
| R25 | HCHO+$NO_2$ = HCO_ν(S)+*trans*-HONO | 0 | 29.6 | 6.7 | 9.2 |
| R26 | HCHO+$NO_2$ = HCO_ν(S)+$HNO_2$ | 0 | 23.3 | 15.8 | 18.3 |
| R27 | HCHO+$NO_2$ = HCO_ν(S)+*cis*-HONO | 0 | 21.2 | 7.6 | 9.6 |
| R28 | $CH_3CHO$ +$NO_2$ = CH3CO_ν(T)+*trans*-HONO | 0 | 28.2 | 6.1 | 10.1 |
| R29 | $CH_3CHO$+$NO_2$ = CH3CO_ν(T)+$HNO_2$ | 0 | 20.7 | 15.0 | 19.2 |
| R30 | $CH_3CHO$+$NO_2$ = CH3CO_ν(T)+*cis*-HONO | 0 | 18.4 | 7.0 | 10.5 |
| R31 | $CH_3CHO$+$NO_2$ = $CH_2CHO$_α(P)+*trans*-HONO | 0 | 54.5 | 8.6 | 16.4 |
| R32 | $CH_3CHO$+$NO_2$ = $CH_2CHO$_α(P)+$HNO_2$ | 0 | 31.0 | 16.7 | 25.5 |
| R33 | $CH_3CHO$+$NO_2$ = $CH_2CHO$_α(P)+*cis*-HONO | 0 | 29.5 | 9.6 | 12.7 |
| | Ethers + $NO_2$ | | | | |
| R34 | $CH_3OCH_3$+$NO_2$ = $CH_3OCH_2$_α(P)+*trans*-HONO | 0 | 31.2 | 12.6 | 17.2 |
| R35 | $CH_3OCH_3$+$NO_2$ = $CH_3OCH_2$_α(P)+$HNO_2$ | 0 | 24.6 | 20.6 | 26.2 |
| R36 | $CH_3OCH_3$+$NO_2$ = $CH_3OCH_2$_α(P)+*cis*-HONO | 0 | 23.3 | 13.6 | 17.5 |
| R37 | $C_2H_5OCH_3$+$NO_2$ = $C_2H_4OCH_3$_β(P)+*trans*-HONO | 0 | 34.5 | 15.2 | 24.2 |
| R38 | $C_2H_5OCH_3$+$NO_2$ = $C_2H_4OCH_3$_β(P)+$HNO_2$ | 0 | 30.5 | 24.4 | 33.2 |
| R39 | $C_2H_5OCH_3$+$NO_2$ = $C_2H_4OCH_3$_β(P)+*cis*-HONO | 0 | 29.1 | 16.8 | 24.5 |
| R40 | $C_2H_5OCH_3$+$NO_2$ = $C_2H_4OCH_3$_α(S)+*trans*-HONO | 0 | 28.9 | 10.8 | 15.9 |
| R41 | $C_2H_5OCH_3$+$NO_2$ = $C_2H_4OCH_3$_α(S)+$HNO_2$ | 0 | 21.7 | 18.8 | 24.9 |
| R42 | $C_2H_5OCH_3$+$NO_2$ = $C_2H_4OCH_3$_α(S)+*cis*-HONO | 0 | 21.5 | 12.1 | 16.2 |
| R43 | $C_3H_7OCH_3$+$NO_2$ = $C_3H_6OCH_3$_α(T)+*trans*-HONO | 0 | 25.9 | 8.2 | 13.9 |
| R44 | $C_3H_7OCH_3$+$NO_2$ = $C_3H_6OCH_3$_α(T)+$HNO_2$ | 0 | 18.5 | 15.8 | 23.0 |
| R45 | $C_3H_7OCH_3$+$NO_2$ = $C_3H_6OCH_3$_α(T)+*cis*-HONO | 0 | 17.1 | 8.3 | 14.2 |

The reactive energies of H-atom abstractions from alcohols, aldehydes and ethers are detailed in Table 1, where product complexes are identified for all reactions. And the Potential energy surface (PES) for H-atom abstraction by $NO_2$ from alcohols, aldehydes and ethers to form $HNO_2$ isomers and the relative product can be seen in Figs. S1 – S3.

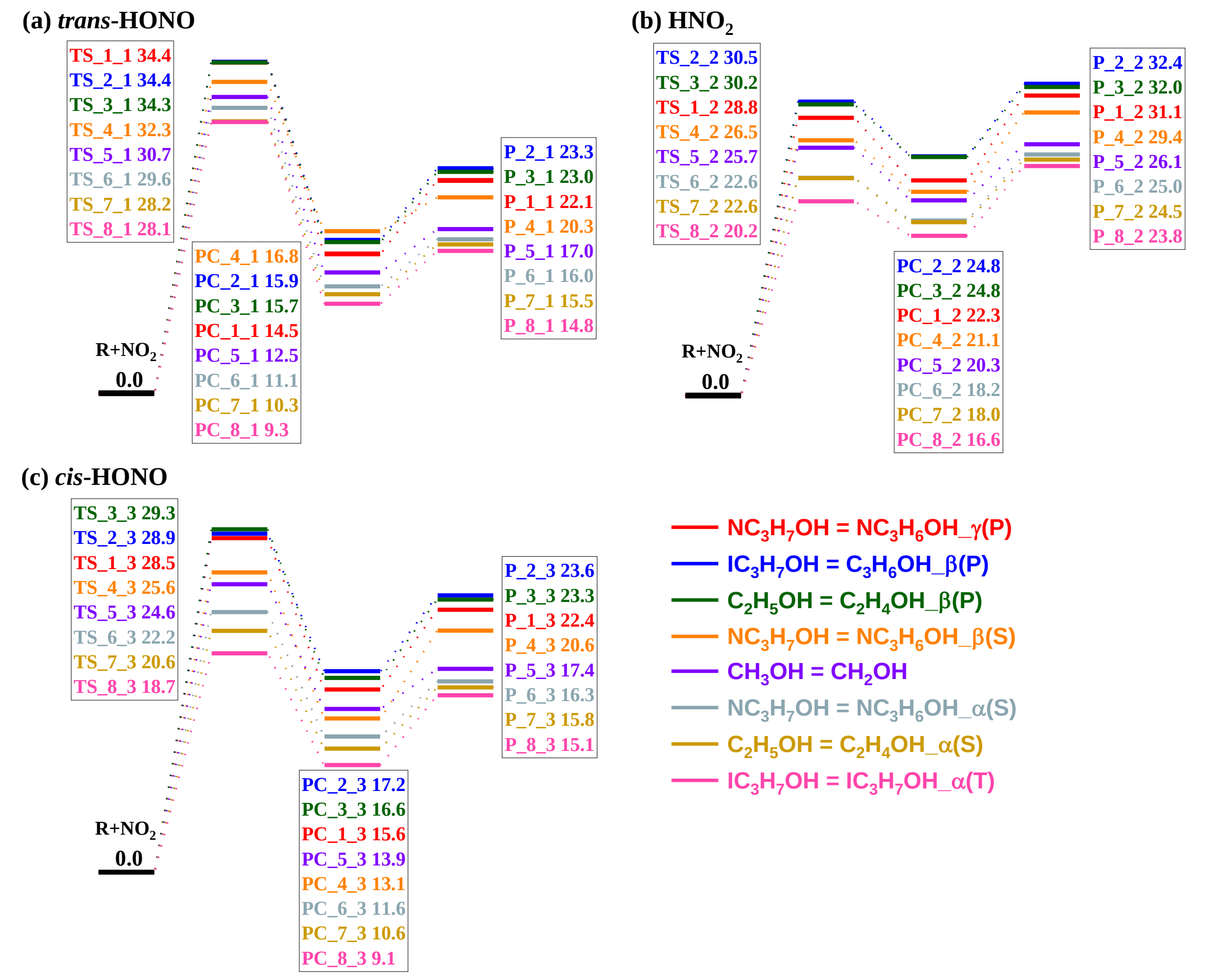

Figure 3. The PES for H-atom abstraction by $NO_2$ from alcohols to form (a)*trans*-HONO, (b) $HNO_2$, and (c) *cis*-HONO. TS – transition state; PC – product complex; P – product.

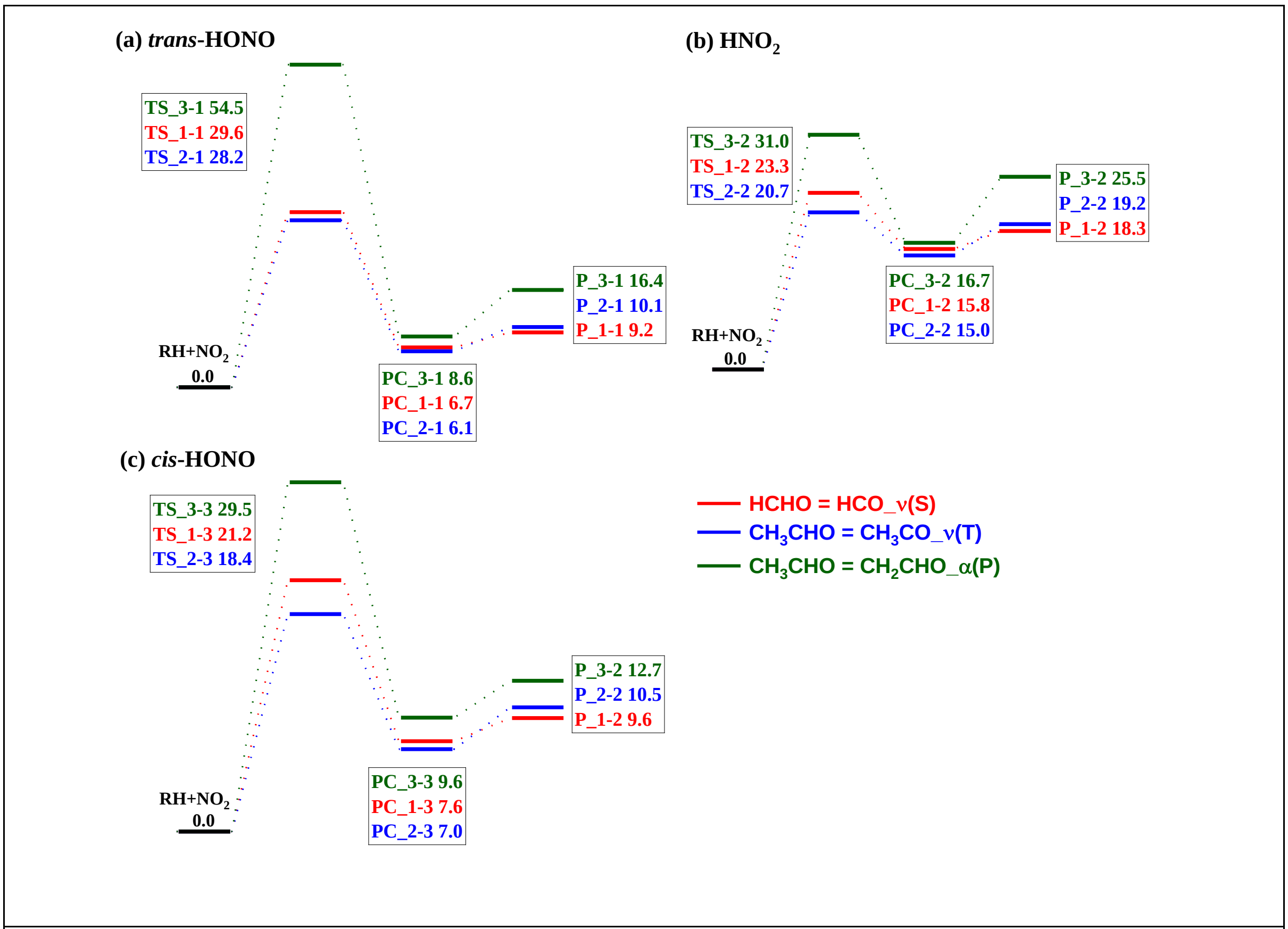

Figure 4. The PES for H-atom abstraction by NO2 from aldehydes to form (a)*trans*-HONO, (b) HNO2, and (c) *cis*-HONO. TS – transition state; PC – product complex; P – product.

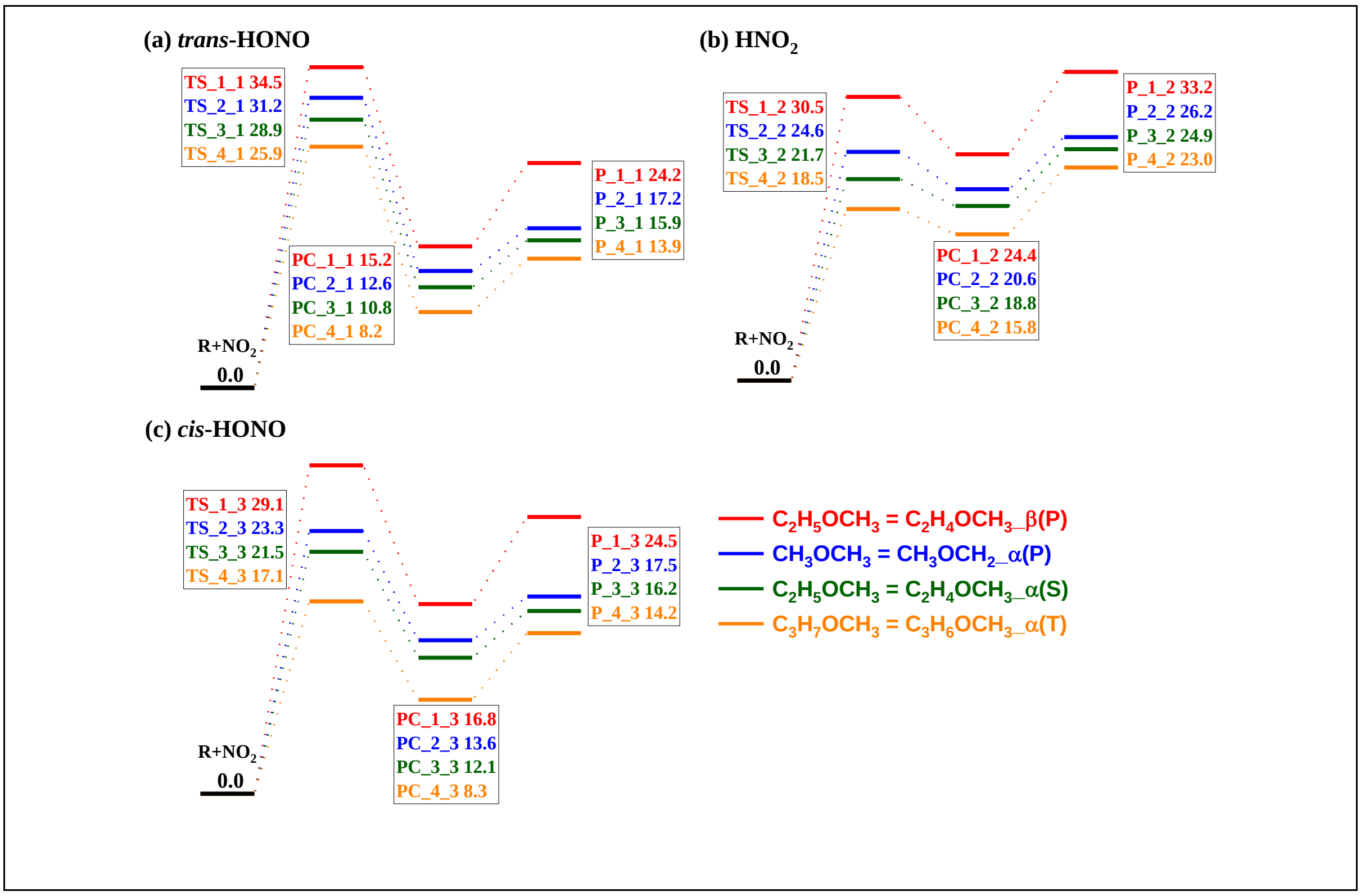

Figure 5. The PES for H-atom abstraction by $NO_2$ from ethers to form (a)*trans*-HONO, (b) $HNO_2$, and (c) *cis*-HONO. TS – transition state; PC – product complex; P – product.

The potential energy surfaces (PESs) for H-atom abstractions by $NO_2$ from alcohols, aldehydes, and ethers are illustracted in Figs. 3-5, respectively. These figures also include all related PCs that connect the TSs with the Ps. For both alcohols and ethers, the energy barriers at various sites leading to the formation of the same $HNO_2$ isomers (*trans*-HONO, $HNO_2$, and *cis*-HONO) follow a similar ranking. In alcohols, the barriers are ranked as β(P) > γ(P) > β(S) > α(P) > α(S) > α(T), Fig.3, while in ethers, the order is β(P) > α(P) > α(S) > α(T), Fig.5. These trends correspond closely to the BDEs of the corresponding sites, as shown in Figs. 1 and 2. It is noteworthy that the energy barriers at the β(P) site are close to those at the γ(P) site, which aligns with the similarity in BDEs at these positions, Figs.1 and 2. In aldehydes (Fig.4), the relative energies of TSs at the ν site are substantially lower than that at the α site. For example, the potential energy required to form $HNO_2$ at the α(P) site is 31.0 kcal/mol, compared to 23.3 kcal/mol at the ν(S) site and 20.7 kcal/mol at the ν(T) site. This disparity is consistent with the significant difference in BDEs between these sites, which is approximately 6 kcal/mol, Fig.1. And the energy barrier of $CH_3CHO$+$NO_2$ = $CH_2CHO$_α(P)+ *trans*-HONO is up to 54.5 kcal/mol, which is the highest one among these reactions. Notably, across all reactions examined, the energy barrier to yield the *trans*-HONO isomer is consistently higher than for the formation of the other two isomers, underscoring its relative difficulty in formation. Mebel et al. [42] attributed this phenomenon to the unique interaction between $NO_2$ and the H atom to form *trans*-HONO, during which the ON π bond on $NO_2$ is disrupted, and the unpaired electron is shifted to the oxygen atom to participate in the formation of the O···H bond. This rearrangement increases repulsive interactions between $NO_2$ and the fuel molecule, thereby elevating the energy barrier for *trans*-HONO formation. Similarly, Chai et al. [43] suggested that the high energy barrier for the *trans*-HONO pathway primarily stems from the need for the RH fragment to approach $NO_2$ more closely. This is necessary to achieve sufficient orbital overlap without cancellation, which in turn leads to heightened Coulombic repulsion.

### *3.3. Rate constant results*

#### *3.3.1. Rate coefficient and uncertainty analysis*

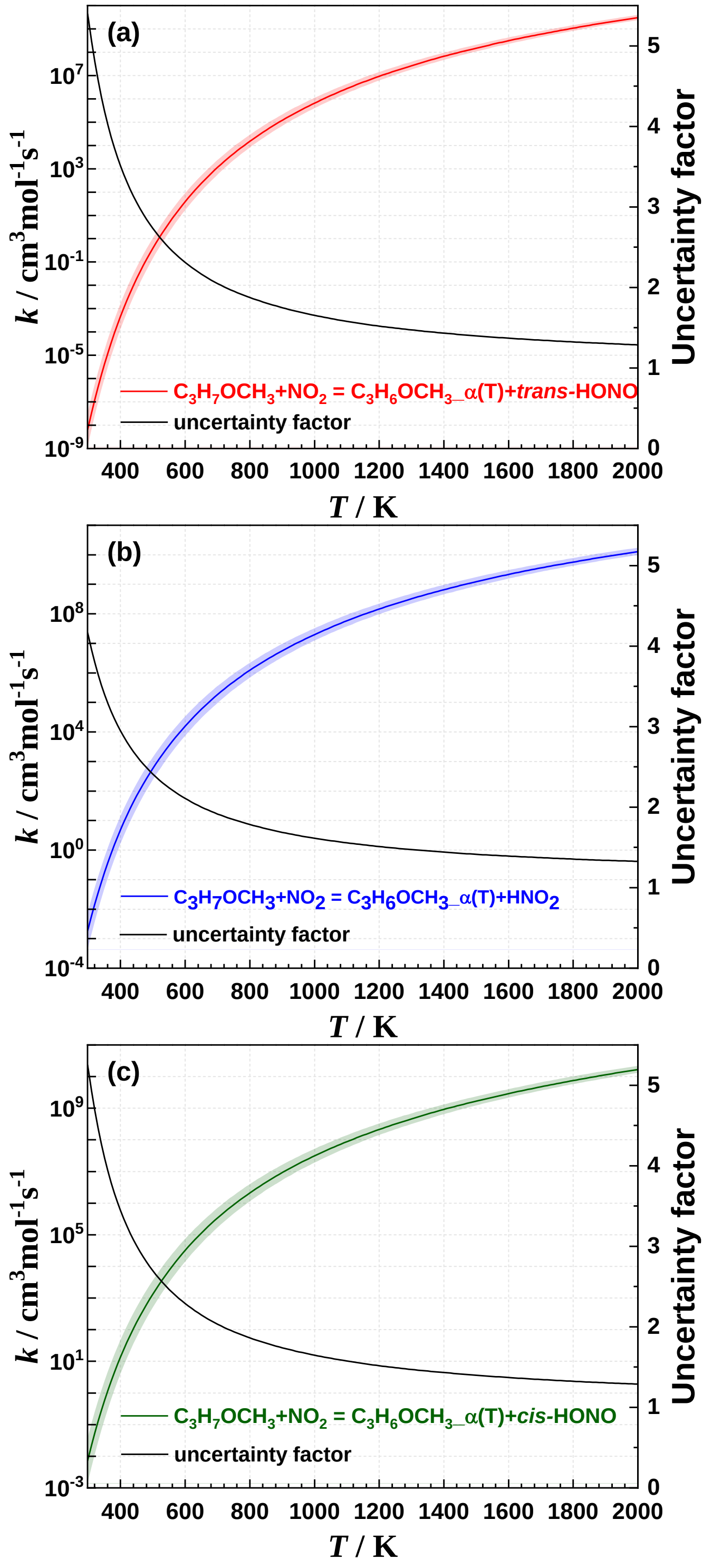

Figure 6. Span of rate coefficients and the respective uncertainty factor for $C_3H_7OCH_3$+$NO_2$ reactions with ±1 kcal/mol change in barrier height.

To quantify the uncertainty in the computed rate coefficients, the well depths of the TSs for the RH + $NO_2$ reactions for $C_3H_7OCH_3$ are adjusted by ±1 kcal/mol, and the rate coefficients are recomputed. An uncertainty factor, *f*, is further defined based on the recomputed rate coefficients, which is determined as $f = \sqrt{\frac{k_{max}}{k_{min}}}$ where $k_{max}$ and $k_{min}$ are the recomputed rate coefficients with the well depth of TS adjusted by -1 kcal/mol and +1 kcal/mol, respectively [44]. $C_3H_7OCH_3$ is selected as the barrier height for these species is the lowest among alcohols, aldehydes and ethers, and applying the same level of absolute energy change to its barrier height will introduce the largest relative change, hence greater impact on the respective rate coefficients. This can serve as an extreme case to demonstrate the uncertainties of the computed rate coefficients. The results are illustrated in Fig. 6. As can be seen from Figs. 6(a) - 6(c), the uncertainty factors for the rate coefficients vary between 1.3 and 5.4 across all three pathways when the barrier height is modified by ±1 kcal/mol. It is noted that the uncertainty is more pronounced at lower temperatures compared to higher temperatures. Note that the uncertainties in Fig. 6 only represent the uncertainties originated from energy calculations, whereas other sources of uncertainties also exist (e.g., 1-D rotor treatment). Nevertheless, a comprehensive evaluation of the uncertainty in rate coefficient computation is beyond the scope of this study, and the uncertainty in barrier height is expected to impose the greatest impact on the computed rate coefficient.

The rate coefficients for H-atom abstraction by $NO_2$, fitted into the Arrhenius expression, are summarized in Table 2. The rate coefficient of same reactants forming different products (*trans*-HONO, $HNO_2$, and *cis*-HONO), in the temperature range from 298K to 2000K, are illustrated in Figs. S4 - S6 in the supplementary Material. In all cases, the H-atom abstraction reactions to form *trans*-HONO have the lowest rate coefficient among all products at almost all temperatures. Especially, the difference between *trans*-HONO and other products are salient at 298K, with at least 3 orders. And the difference will decrease when the temperature increases.

Table 2 The rate coefficient for H-atom abstraction by $NO_2$ from alcohols, aldehydes and ethers to form the respective products and $HNO_2$ isomers (*trans*-HONO, $HNO_2$, *cis*-HONO).

| No. | Reaction | A ($cm^3$/mol*s) | n | Ea (cal/mol) |
|---|---|---|---|---|
| | Alcohols + $NO_2$ | | | |
| R1 | $CH_3OH$+$NO_2$ = $CH_2OH$+*trans*-HONO | 1.000E+00 | 3.721 | 28711.90 |
| R2 | $CH_3OH$ +$NO_2$ = $CH_2OH$+$HNO_2$ | 1.000E+00 | 3.729 | 22982.83 |
| R3 | $CH_3OH$ +$NO_2$ = $CH_2OH$+*cis*-HONO | 1.000E+00 | 3.878 | 21928.76 |
| R4 | $C_2H_5OH$+$NO_2$ = $C_2H_4OH$_α(S)+*trans*-HONO | 1.000E+00 | 3.591 | 26929.96 |
| R5 | $C_2H_5OH$ +$NO_2$ = $C_2H_4OH$_α(S)+$HNO_2$ | 1.000E+00 | 3.628 | 20301.74 |
| R6 | $C_2H_5OH$ +$NO_2$ = $C_2H_4OH$_α(S)+*cis*-HONO | 1.000E+00 | 3.617 | 19247.99 |
| R7 | $C_2H_5OH$ +$NO_2$ = $C_2H_4OH$_β(P)+*trans*-HONO | 1.000E+00 | 3.819 | 33327.35 |
| R8 | $C_2H_5OH$ +$NO_2$ = $C_2H_4OH$_β(P)+$HNO_2$ | 1.000E+00 | 3.647 | 29199.90 |
| R9 | $C_2H_5OH$ +$NO_2$ = $C_2H_4OH$_β(P)+*cis*-HONO | 1.000E+00 | 3.730 | 26523.37 |
| R10 | $NC_3H_7OH$+$NO_2$ = $NC_3H_6OH$_α(S)+*trans*-HONO | 1.000E+00 | 3.764 | 28390.75 |
| R11 | $NC_3H_7OH$+$NO_2$ = NC3H6OH_α(S)+$HNO_2$ | 1.000E+00 | 3.579 | 20445.37 |
| R12 | $NC_3H_7OH$+$NO_2$ = NC3H6OH_α(S)+*cis*-HONO | 1.000E+00 | 3.674 | 20494.01 |
| R13 | $NC_3H_7OH$+$NO_2$ = NC3H6OH_β(S)+*trans*-HONO | 5.369E+01 | 3.237 | 31304.19 |
| R14 | $NC_3H_7OH$+$NO_2$ = NC3H6OH_β(S)+$HNO_2$ | 1.000E+00 | 3.659 | 25086.21 |
| R15 | $NC_3H_7OH$+$NO_2$ = NC3H6OH_β(S)+*cis*-HONO | 1.000E+00 | 3.729 | 23276.18 |
| R16 | $NC_3H_7OH$+$NO_2$ = $NC_3H_6OH$_γ(P)+*trans*-HONO | 6.429E+00 | 3.704 | 32984.41 |
| R17 | $NC_3H_7OH$+$NO_2$ = $NC_3H_6OH$_γ(P)+$HNO_2$ | 1.000E+00 | 3.793 | 26472.59 |
| R18 | $NC_3H_7OH$+$NO_2$ = $NC_3H_6OH$_γ(P)+*cis*-HONO | 1.000E+00 | 3.862 | 25780.16 |
| R19 | $IC_3H_7OH$+$NO_2$ = $IC_3H_6OH$_α(T)+ *trans*-HONO | 1.000E+00 | 3.691 | 26639.03 |
| R20 | $IC_3H_7OH$ +$NO_2$ = $IC_3H_6OH$_α(T)+$HNO_2$ | 1.000E+00 | 3.594 | 17916.55 |
| R21 | $IC_3H_7OH$ +$NO_2$ = $IC_3H_6OH$_α(T)+*cis*-HONO | 1.000E+00 | 3.615 | 17374.20 |
| R22 | $IC_3H_7OH$+$NO_2$ = $IC_3H_6OH$_β(P)+*trans*-HONO | 1.000E+00 | 3.888 | 33097.53 |
| R23 | $IC_3H_7OH$ +$NO_2$ = $IC_3H_6OH$_β(P)+$HNO_2$ | 1.000E+00 | 3.695 | 28590.87 |
| R24 | $IC_3H_7OH$ +$NO_2$ = $IC_3H_6OH$_β(P)+*cis*-HONO | 1.000E+00 | 3.762 | 27003.32 |
| | Aldehydes + $NO_2$ | | | |
| R25 | HCHO+$NO_2$ = HCO_ν(S)+*trans*-HONO | 1.001E+00 | 3.804 | 25860.61 |
| R26 | HCHO+$NO_2$ = HCO_ν(S)+$HNO_2$ | 1.001E+00 | 3.849 | 18907.55 |
| R27 | HCHO+$NO_2$ = HCO_ν(S)+*cis*-HONO | 1.001E+00 | 3.777 | 17074.71 |
| R28 | $CH_3CHO$ +$NO_2$ = CH3CO_ν(T)+*trans*-HONO | 1.001E+00 | 3.765 | 24909.60 |
| R29 | $CH_3CHO$+$NO_2$ = CH3CO_ν(T)+$HNO_2$ | 1.001E+00 | 3.703 | 17094.13 |
| R30 | $CH_3CHO$+$NO_2$ = CH3CO_ν(T)+*cis*-HONO | 1.001E+00 | 3.747 | 15350.08 |
| R31 | $CH_3CHO$+$NO_2$ = $CH_2CHO$_α(P)+*trans*-HONO | 1.001E+00 | 3.994 | 51223.04 |
| R32 | $CH_3CHO$+$NO_2$ = $CH_2CHO$_α(P)+$HNO_2$ | 1.001E+00 | 3.832 | 27118.22 |
| R33 | $CH_3CHO$+$NO_2$ = $CH_2CHO$_α(P)+*cis*-HONO | 1.001E+00 | 3.870 | 25667.03 |
| | Ethers + $NO_2$ | | | |
| R34 | $CH_3OCH_3+NO_2 = CH_3OCH_2$_α(P)+*trans*-HONO | 1.001E+00 | 4.013 | 29466.22 |
| R35 | $CH_3OCH_3+NO_2 = CH_3OCH_2$_α(P)+$HNO_2$ | 1.001E+00 | 3.787 | 21543.59 |
| R36 | $CH_3OCH_3+NO_2 = CH_3OCH_2$_α(P)+*cis*-HONO | 1.001E+00 | 3.958 | 20411.24 |
| R37 | $C_2H_5OCH_3+NO_2 = C_2H_4OCH_3$_β(P)+*trans*-HONO | 1.001E+00 | 3.871 | 32780.32 |
| R38 | $C_2H_5OCH_3+NO_2 = C_2H_4OCH_3$_β(P)+$HNO_2$ | 1.001E+00 | 3.651 | 27992.68 |
| R39 | $C_2H_5OCH_3+NO_2 = C_2H_4OCH_3$_β(P)+*cis*-HONO | 1.000E+00 | 3.849 | 25308.67 |
| R40 | $C_2H_5OCH_3+NO_2 = C_2H_4OCH_3$_α(S)+*trans*-HONO | 1.001E+00 | 3.712 | 26608.89 |
| R41 | $C_2H_5OCH_3+NO_2 = C_2H_4OCH_3$_α(S)+$HNO_2$ | 1.001E+00 | 3.685 | 19010.27 |
| R42 | $C_2H_5OCH_3+NO_2 = C_2H_4OCH_3$_α(S)+*cis*-HONO | 1.001E+00 | 3.748 | 18627.81 |
| R43 | $C_3H_7OCH_3+NO_2 = C_3H_6OCH_3$_α(T)+*trans*-HONO | 1.001E+00 | 3.653 | 23510.81 |
| R44 | $C_3H_7OCH_3+NO_2 = C_3H_6OCH_3$_α(T)+$HNO_2$ | 1.001E+00 | 3.587 | 15836.68 |
| R45 | $C_3H_7OCH_3+NO_2 = C_3H_6OCH_3$_α(T)+*cis*-HONO | 1.001E+00 | 3.595 | 15053.00 |

### *3.3.2. Comparison of rate coefficients with other studies*

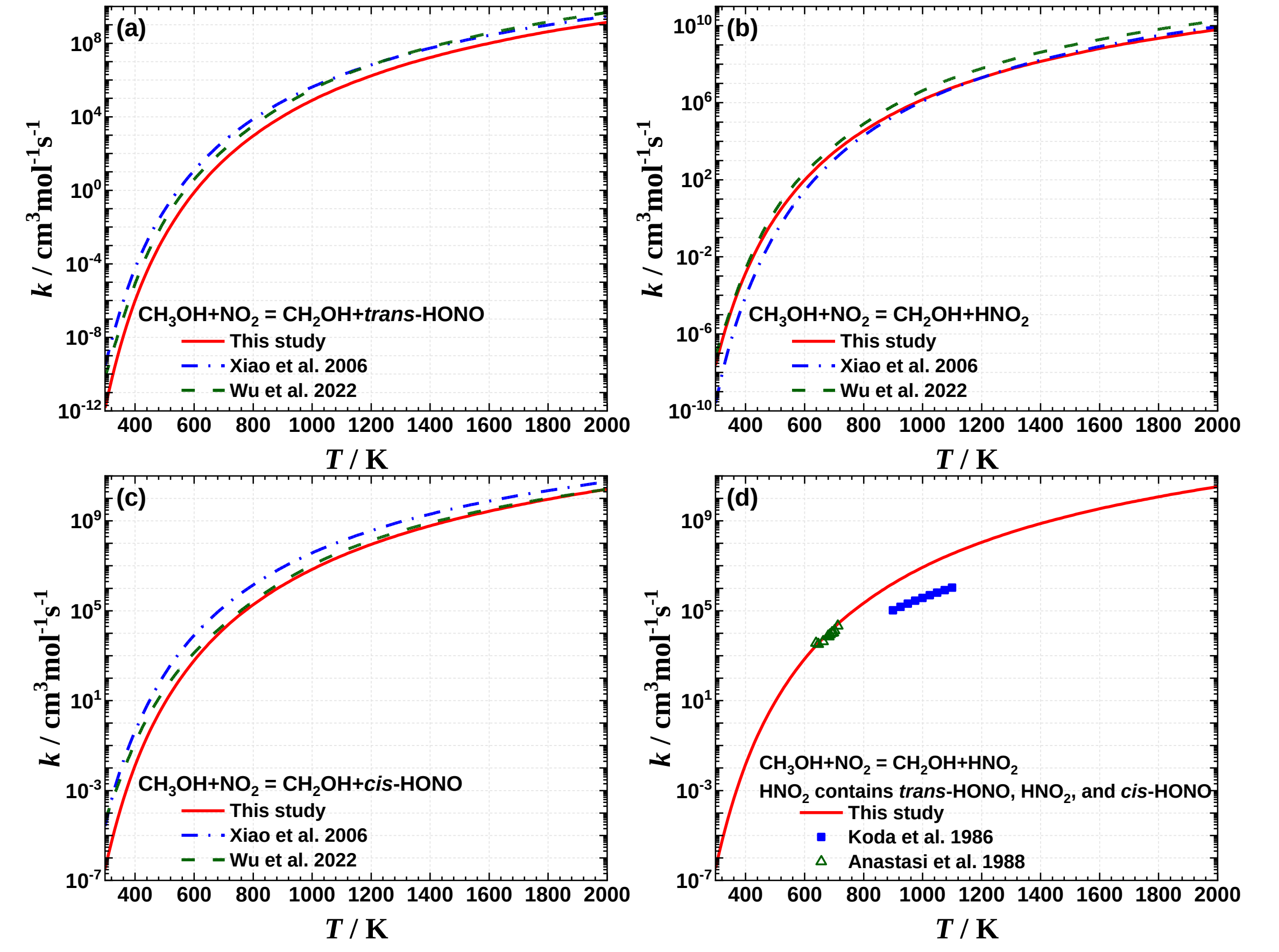

Figure 7. Comparison of rate coefficients for H-atom abstraction by $NO_2$ from $CH_3OH$ to form $CH_2OH$: Results from this study, Xiao et al. [16], Wu et al. [1], experimental data from Koda et al. [18], and Anastasi et al. [17].

To validate the calculations in this work, the calculated rate coefficients are compared with the theoretical computations and experimental measurements from previous studies. Figure 7 summarizes the comparison results for $CH_3OH$ forming $CH_2OH$, where Fig. 7(d) presents the total abstraction rate coefficients as Koda et al. [18] and Anastasi et al. [17] did not distinguish the produced $HNO_2$ isomers during the measurements. It is clear from Figs. 7(a) – 7(c) that the rate coefficients calculated by this study agree well with previous computational studies over the whole temperature range. And for *cis*-HONO, the rate coefficients calculated by this study are little lower than previous computational studies. However, A better agreement is observed with experimental measurements, as shown in Fig. 7(d). The calculated results coincide with the experimental data from Anastasi et al. [17] and have some discrepancies within the experimental measurements from Koda et al. [18]. Computational uncertainty is illustrated in Fig. 6.

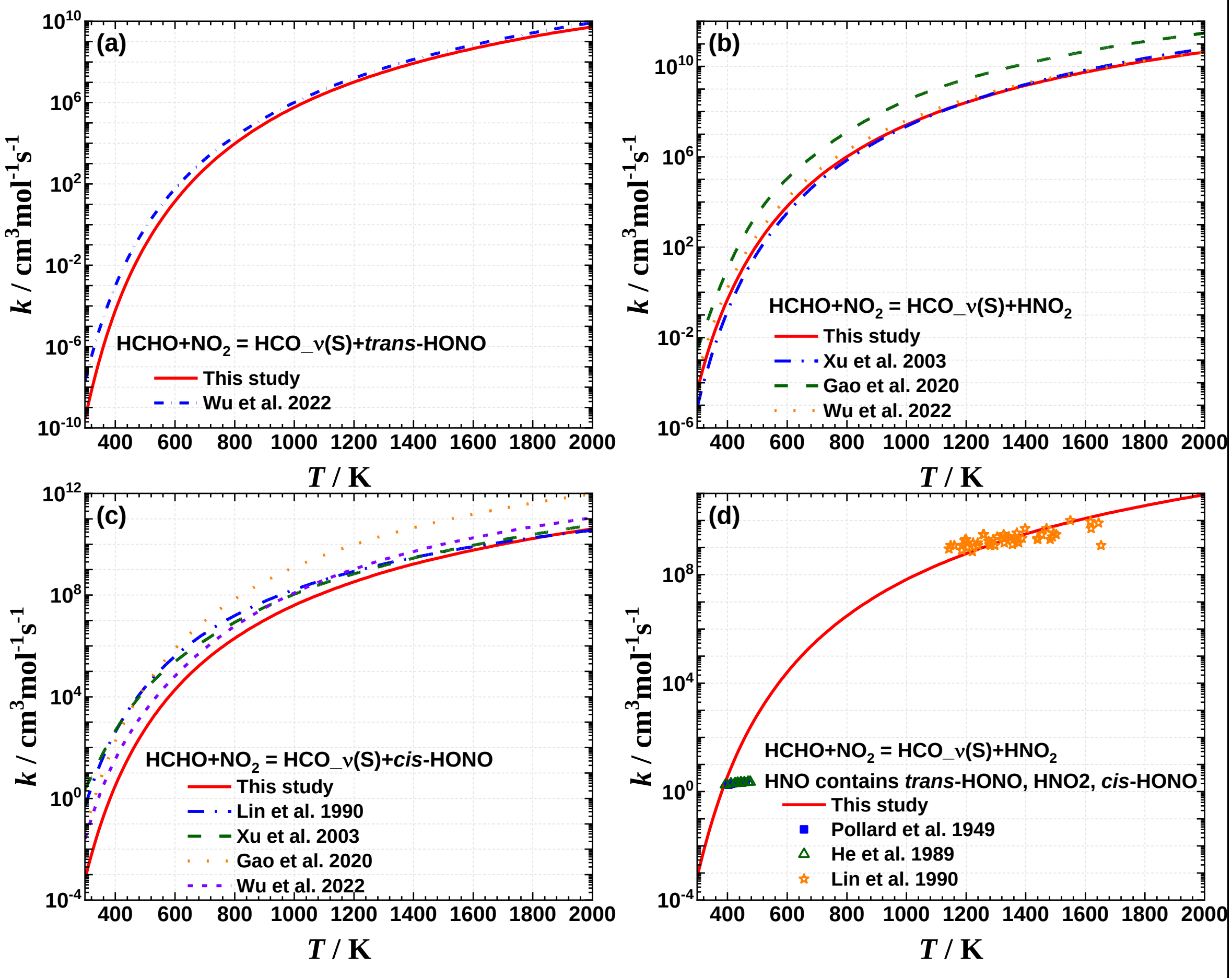

Figure 8. Comparison of rate coefficients for H-atom abstraction by $NO_2$ from HCHO to form HCO_v(S): Results from this study, Lin et al. [21], Xu et al. [22], Gao et al. [23], Wu et al. [1], experimental data from Pollard et al. [19], He et al. [20], and Lin et al. [21].

The comparison for HCHO is further illustrated in Fig.8. Notably, the rate coefficients calculated by this study agree well with previous computational studies over the whole temperature ranges, with slightly better agreements observed with those from Wu et al. [1] where a higher level of theory was used than others. For *cis*-HONO channel, the rate coefficient calculated by this study is lower than other calculations with about 2-3 orders of magnitude. However, the mixed rate coefficient has a better agreement with the experiment data, due to the reason that the major channel of *cis*-HONO have a lower rate coefficient than other theoretical calculations.

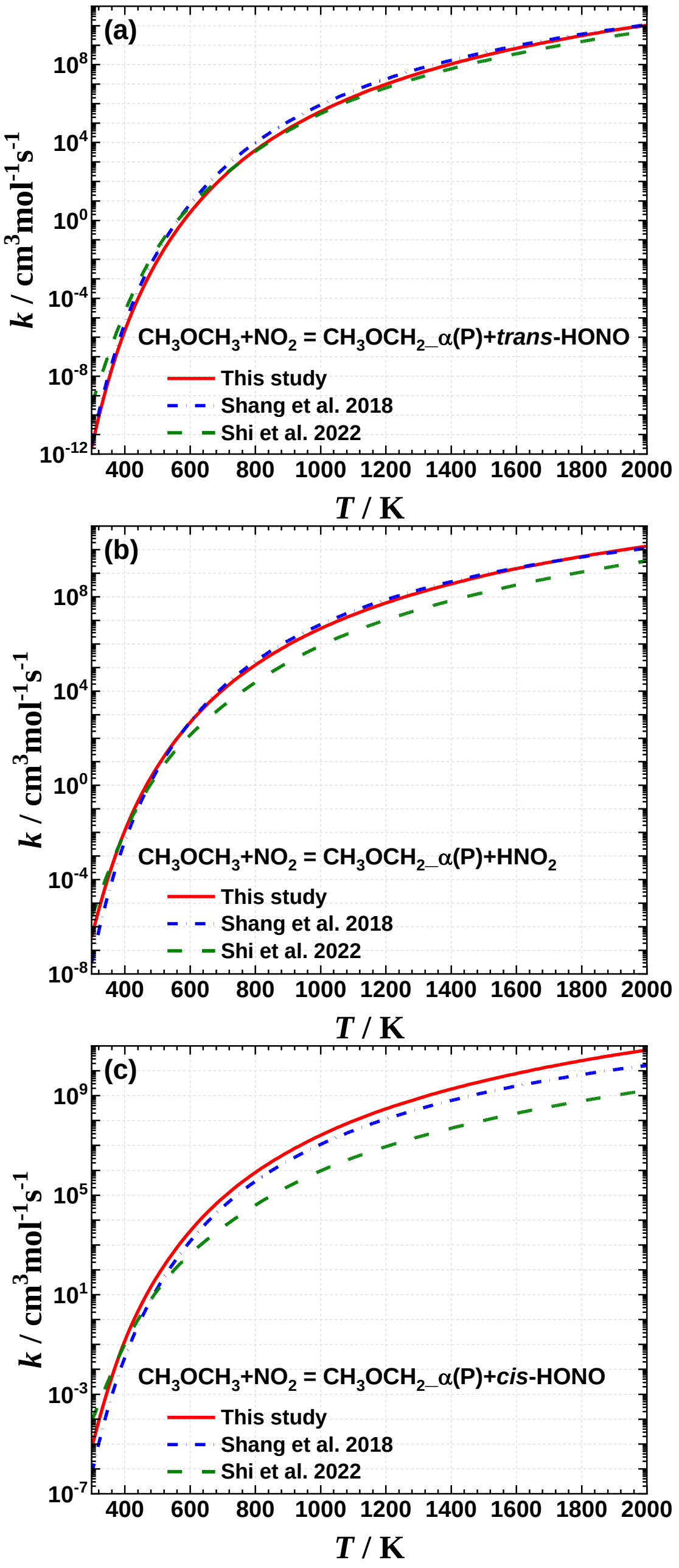

Figure 9. Comparison of rate coefficients for H-atom abstraction by $NO_2$ from $CH_3OCH_3$ to form $CH_3OCH_2$_α(P): Results from this study, Shang et al. [2], and Shi et al. [24].

When it comes to the comparison of $CH_3OCH_3$ rate coefficients in Fig.9, there is, again, the rate coefficients calculated by this investigation show good agreement with those reported by Shang et al. [2] However, there are some differences when compared to the results of Shi et al. [24]. These discrepancies may arise from the fact that Shi et al. [24] did not account for the 1-D hindered rotor effect in their calculations. The differences are generally within an order of magnitude, which can be equivalent to the combined uncertainty from the computations. The comparison results for other ether species are summarized in the Supplementary Material (i.e., Figs. S7 and S8).

### *3.3.3. Comparison of rate coefficient between different species at the same site*

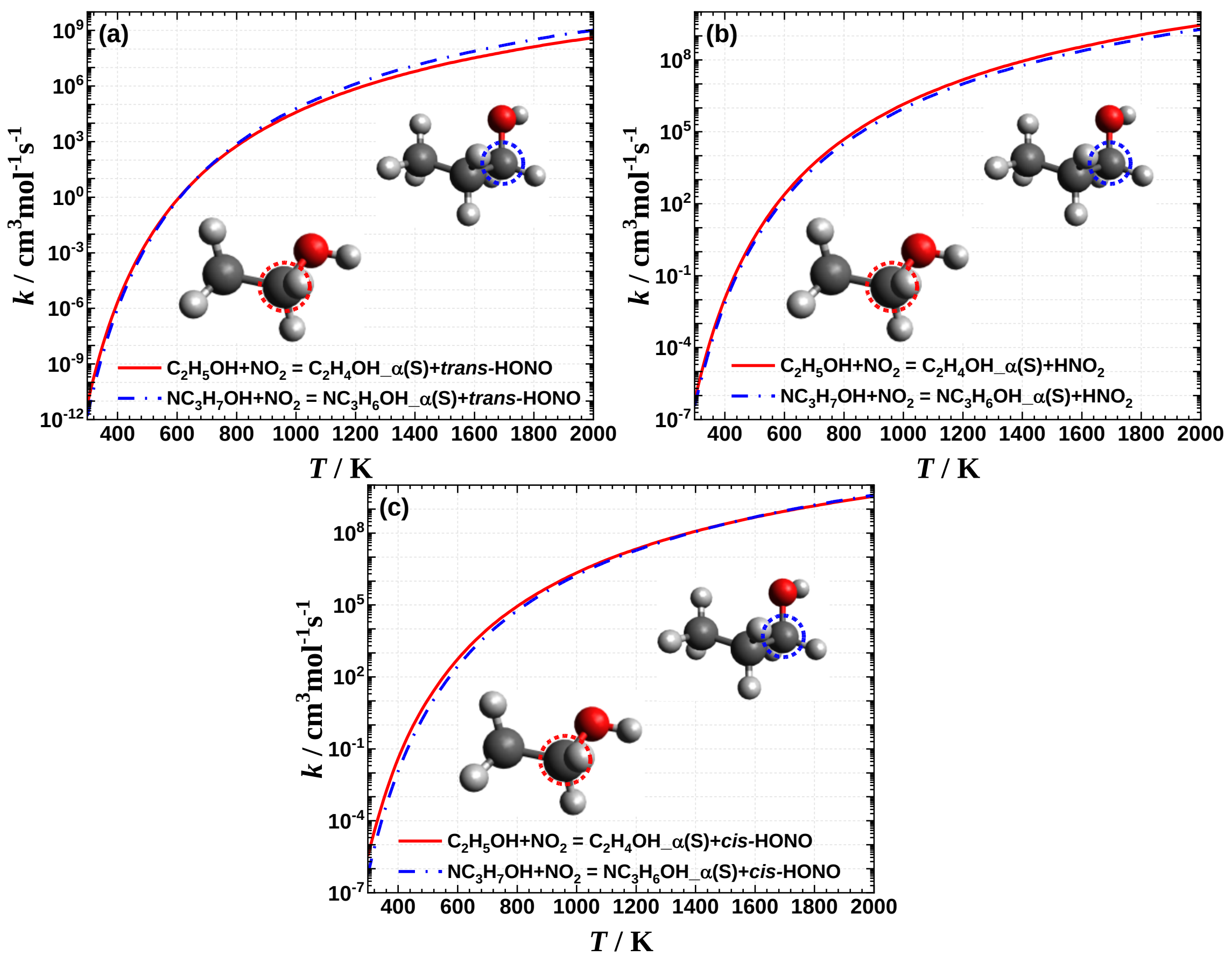

Figure 10. The comparison between the rate coefficients for H-atom abstraction by $NO_2$ at α(S) site from $C_2H_5OH$ and $NC_3H_7OH$ to form (a) *trans*-HONO, (b) $HNO_2$, and (c) *cis*-HONO. Data is presented on per H atom basis.

Figure 10 presents the comparison of the rate coefficients at the α(S) site of $C_2H_5OH$, and $NC_3H_7OH$. The rate coefficients for these alcohols are nearly identical across the entire temperature range, indicating that the number of carbon atoms has little impact on the coefficient rate at this site. The level of difference is within the calculation uncertainty as discussed in Fig. 6. The comparisons of rate coefficients between different species at the same site are presented in Fig. S9, where similar trends are observed.

*3.3.4. Comparison of rate coefficient between different sites*

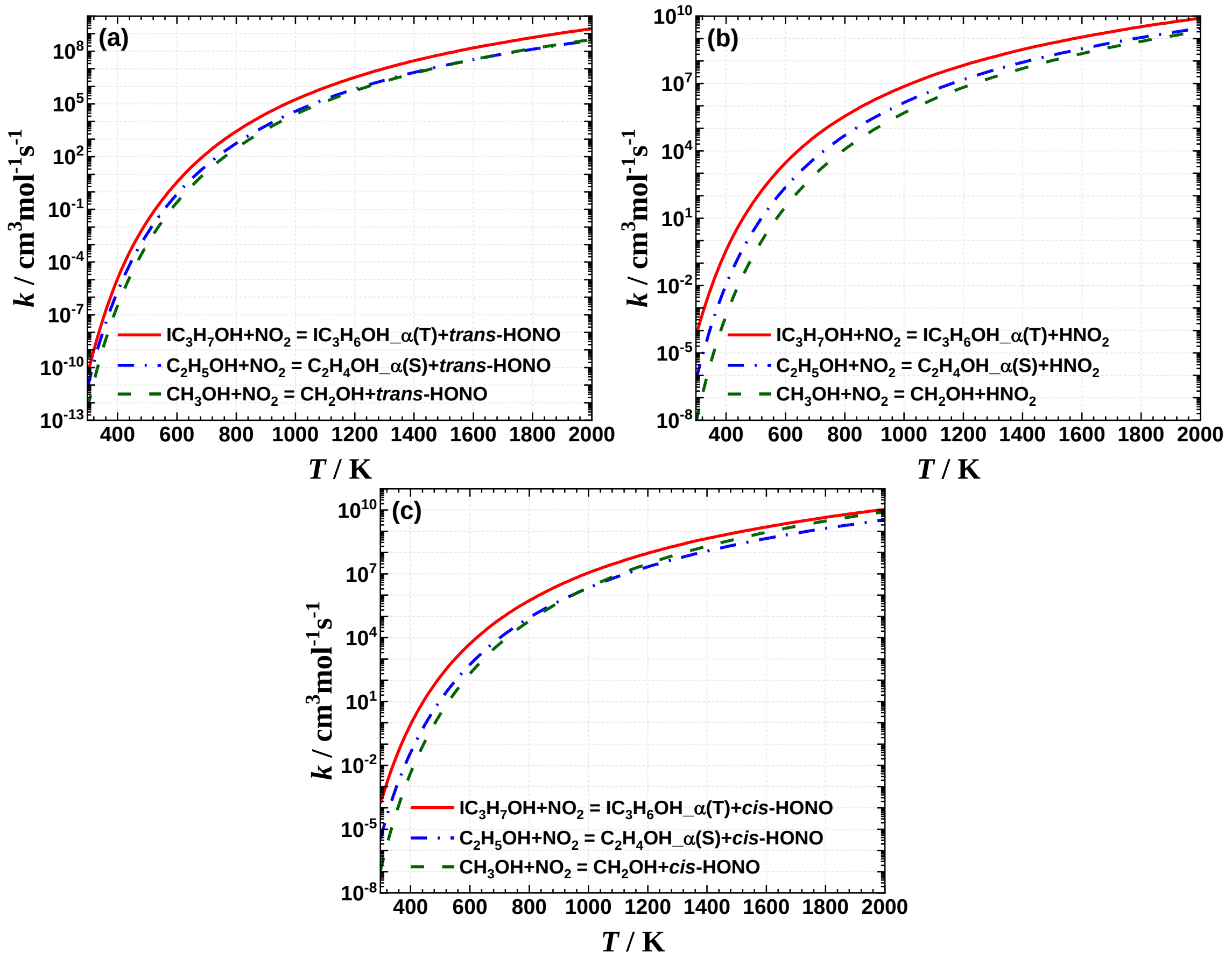

Figure 11. The comparison between the rate coefficients for H-atom abstraction by $NO_2$ at α(T) site from $IC_3H_7OH$, α(S) site from $C_2H_5OH$, and α(P) site from $CH_3OH$ to form (a) *trans*-HONO, (b) $HNO_2$, and (c) *cis*-HONO. Data is presented on per H atom basis.

Figure 11 compares the H-atom abstractions at the α(T), α(S), and α(P) site of alcohols. It should be noted that the rate coefficient at the α(T) site is consistently the highest H-atom reaction, follow with α(S) site and α(P) site, Figs. 11(a) – 11(c). This distribution is consistent with the energy barriers at the α(P), α(S), and α(T) sites as shown in Fig. 3. As the temperature increases, the difference in rate coefficients among all sites diminishes, eventually narrowing to 1 order of magnitude at 2000 K.

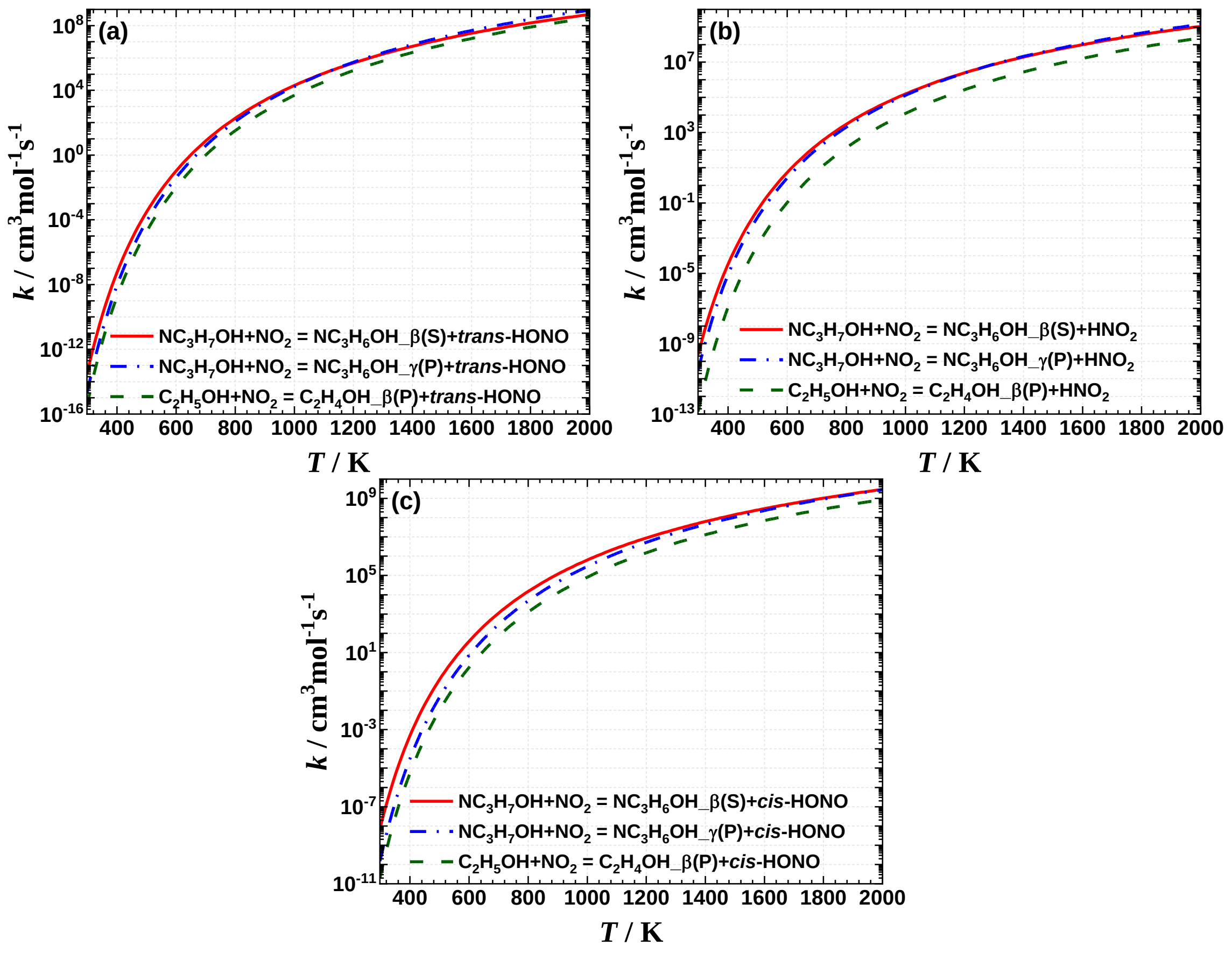

Figure 12. The comparison between the rate coefficients for H-atom abstraction by $NO_2$ at β(S) site from $NC_3H_7OH$, γ(P) site from $NC_3H_7OH$, and β(P) site from $C_2H_5OH$ to form (a) *trans*-HONO, (b) $HNO_2$, and (c) *cis*-HONO. Data is presented on per H atom basis.

Figure 12 presents a comparison of the rate coefficients for H-atom abstraction by $NO_2$ from β(S) site of $NC_3H_7OH$, γ(P) site of $NC_3H_7OH$, and β(P) site of $C_2H_5OH$. The rate coefficient at the β(S) site continuesly exceeds those at the γ(P) and β(P) sites. It should be noted that as the carbon atom is positioned further from the functional group, the influence of the functional group diminishes. For instance, the rate coefficient at the γ(P) site is slightly higher than that of β(P) site in both the *trans*-HONO and *cis*-HONO channels, while it is close to that of the β(S) site in the $HNO_2$ channel. Conversely, the rate coefficient at the α site is significantly higher than those at the other sites at low temperature, Figs. 11 and 12.

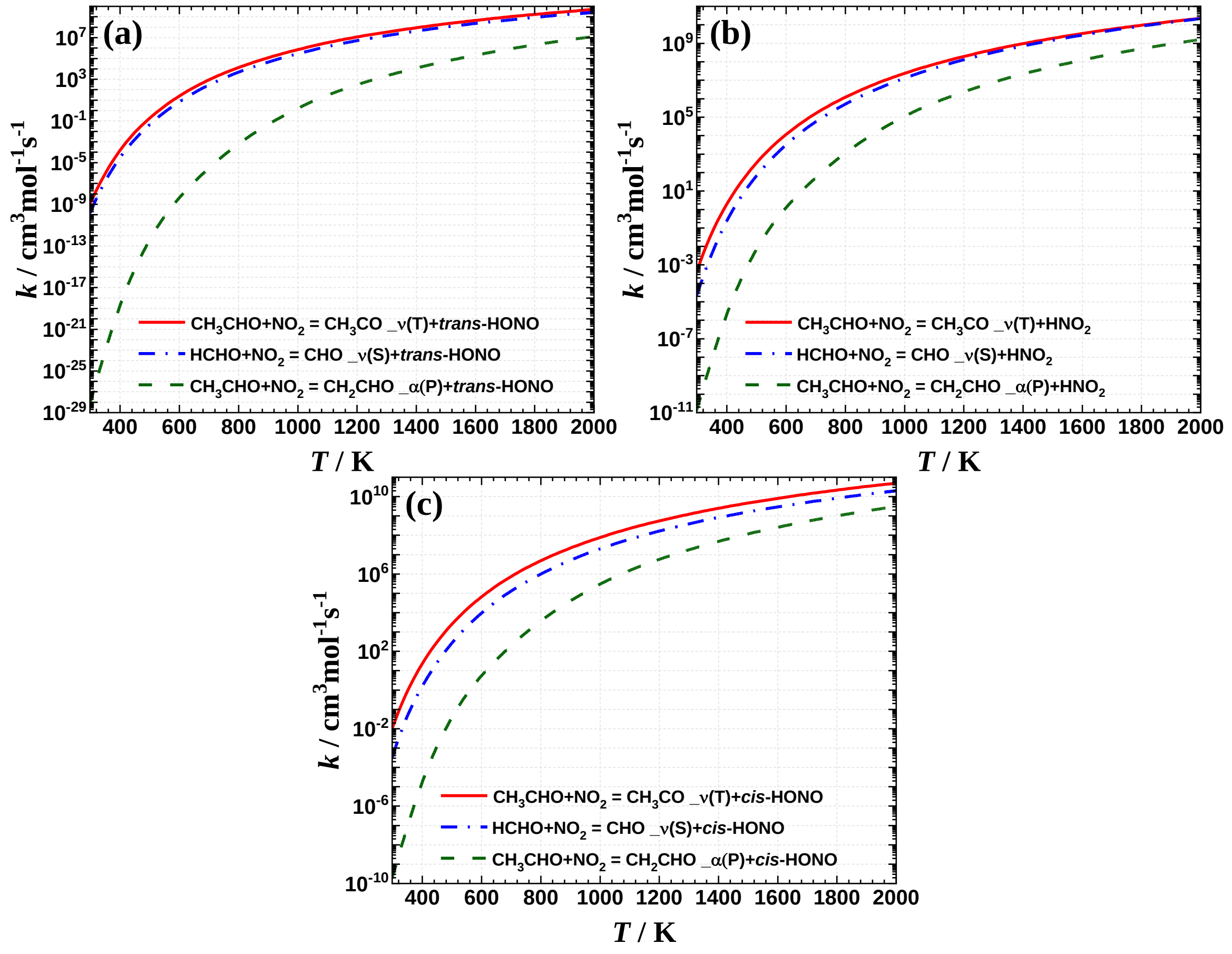

Figure 13. The comparison between the rate coefficients for H-atom abstraction by $NO_2$ at ν(T) site from $CH_3CHO$, ν(S) site from HCHO, and α(P) site from $CH_3CHO$ to form (a) *trans*-HONO, (b) $HNO_2$, and (c) *cis*-HONO. Data is presented on per H atom basis.

Figure 13 explores the difference of rate coefficient by H atom abstraction at different sites from aldehydes, it is clear that the trend of rate coefficient for these kinds of reaction have consistent rank at whole temperature, from high to low: ν(T) > ν(S) > α(P), which links to the energy barriers of corresponding reactions. And the difference between the ν(T) site and the ν(S) site is near 1 order. However, an obvious difference can be found from the α(P) site with the ν(S) site at low temperature, with about 6-18 orders, indicating that the reaction: $CH_3CHO+NO_2$ = $CH_2CHO$_α(P)+*trans*-HONO/$HNO_2$/*cis*-HONO is hard to happen in this reaction system compared to other reactions at low temperature. Then, the rate coefficient at the α(P) has experienced a dramatic increase with temperature, and the difference of both α(P) site and ν(S) site decreases to about 2 orders at 2000K, underscoring that the enhanced influences of hindered rotor effects at higher temperature.

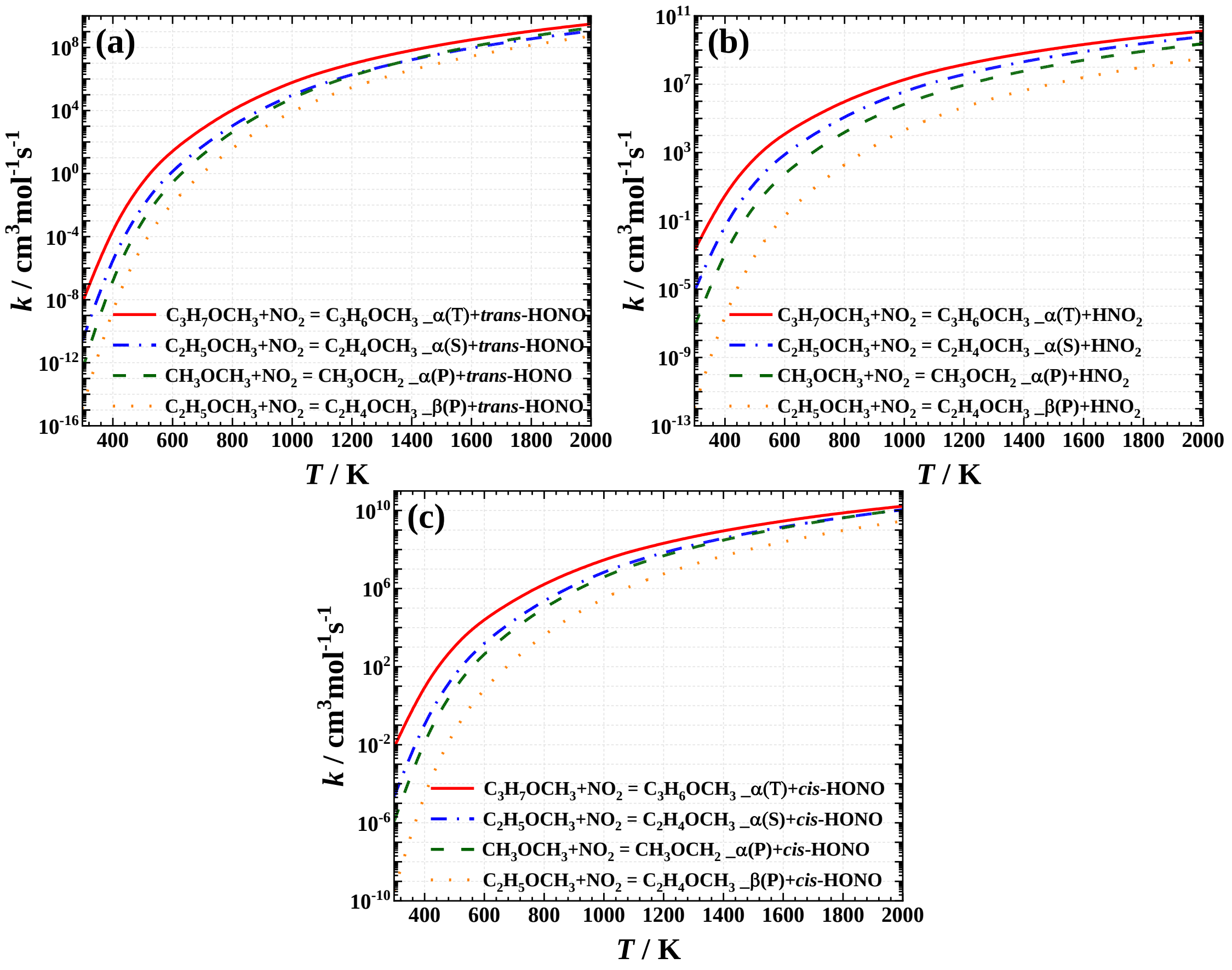

Figure 14. The comparison between the rate coefficients for H-atom abstraction by $NO_2$ at the α(T) site from $C_3H_7OCH_3$, α(S) site from $C_2H_5OCH_3$, α(P) site from $CH_3OCH_3$, and β(P) site from $C_2H_5OCH_3$ to form (a) *trans*-HONO, (b) $HNO_2$, and (c) *cis*-HONO. Data is presented on per H atom basis.

Figure 14 focuses on comparing the rate coefficients for H-atom abstraction by the $NO_2$ from ethers. Within the temperature range of 298 K to 2000 K, the trend of these rate coefficients ranks from high to low: α(T) > α(S) > α(P) > β(P). Particularly noteworthy is the rate coefficient of α(P) site rapid increases to exceed that of α(S) site with increasing temperature, Figs. 14(a) and 14(c). The reason is related to the hindered rotor promoting the reaction rate.

As illustrated in Figs. 11-14, the discrepancies in the rate coefficients for H-atom abstraction at different sites to form $HNO_2$ isomers are most pronounced at 298K, with the differences consistently decreasing as the temperature increases. As for the different sites at identical reactants, the comparison between the rate coefficients for H-atom abstraction to form $HNO_2$ isomers has also been conducted for $C_2H_5OH$, $NC_3H_7OH$, $IC_3H_7OH$, $CH_3CHO$, and $C_2H_5OCH_3$ (as can be seen in Figs. S10-S14).

### *3.4. Branching ratio analysis*

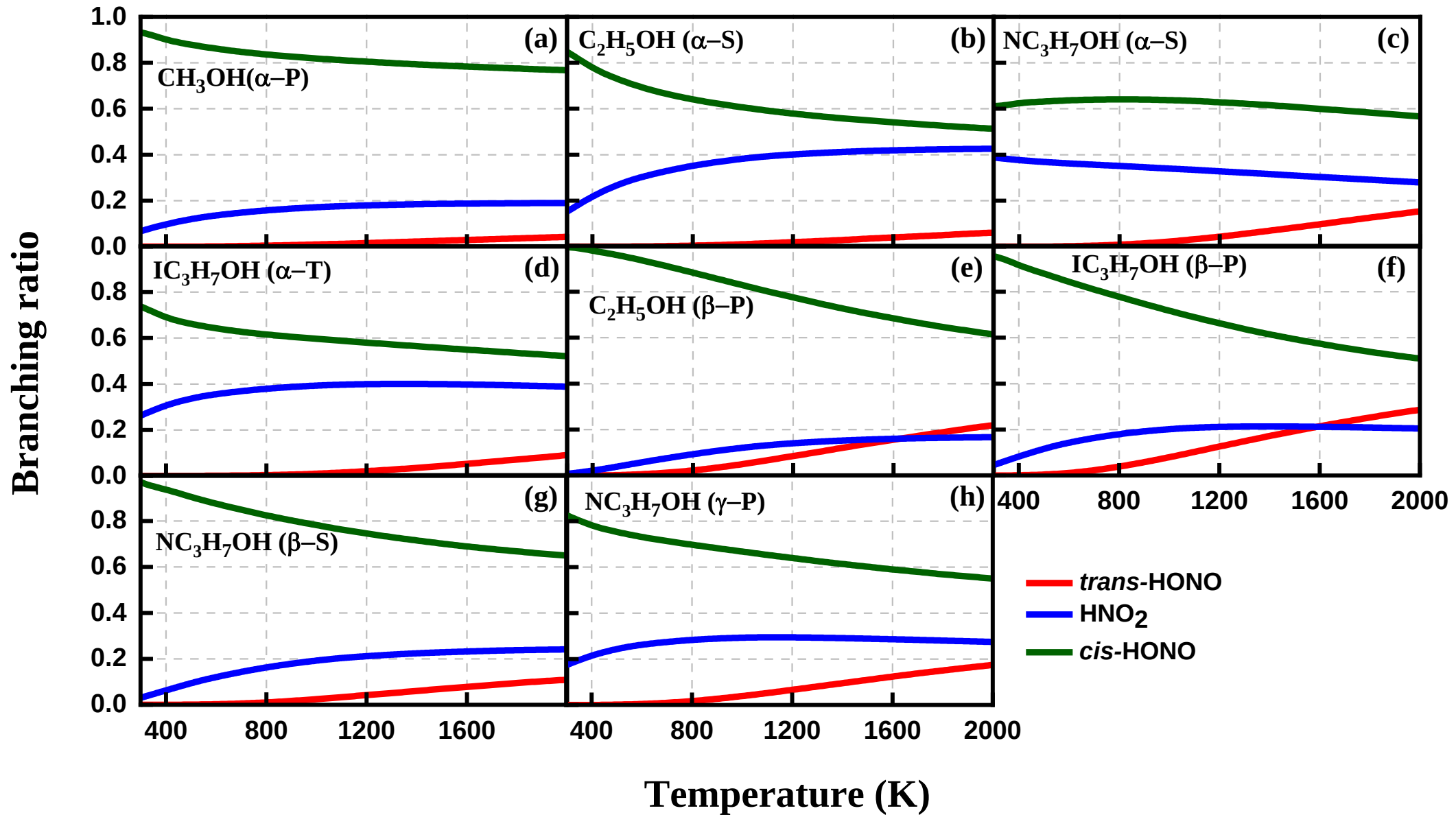

Figure 15. The comparison between the branching ratio for H-atom abstraction by $NO_2$ from alcohols to form $HNO_2$ isomers (*trans*-HONO, $HNO_2$, *cis*-HONO), (a) $CH_3OH$ (α-P), (b) $C_2H_5OH$ (α-S), (c) $NC_3H_7OH$ (α-S), (d) $IC_3H_7OH$ (α-T), (e) $C_2H_5OH$ (β-P), (f) $IC_3H_7OH$ (β-P), (g) $NC_3H_7OH$ (β-S), (h) $NC_3H_7OH$ (γ-P).

Figure 15 compares the branching ratio for H-atom abstraction by $NO_2$ from alcohols. Totally, compared with other two channels, the channel of *cis*-HONO occupies the dominancy ratio throughout the entire temperature range. Except $NC_3H_7OH$ (α-S), the proportion of *cis*-HONO channel gradually decreases with the decreasing temperature, following with the enhance of *trans*-HONO and $HNO_2$ channel. For $NC_3H_7OH$ (α-S), Fig.15(c), the *cis*-HONO channel stands at about 60% at 298K and has a slight increase until about 800K, then gradually decreases to around 48% at 2000K. In the contract, the channel of $HNO_2$ has a consistently decline with the temperature range from 298K to 2000K. This different trend is due to the reason that the rate coefficient of H atom abstraction at (α-S) site by $NC_3H_7OH$ to form *cis*-HONO is approach to that of $HNO_2$ between the temperature range of 298K to about 800K, the difference between them is within 1 order, Fig. S4(d). As the temperature increases, the difference gradually widens. Across all species, the proportion of *trans*-HONO remains relatively low, typically not exceeding 25%, indicating that this pathway is less favored compared to the pathways forming *cis*-HONO and $HNO_2$.

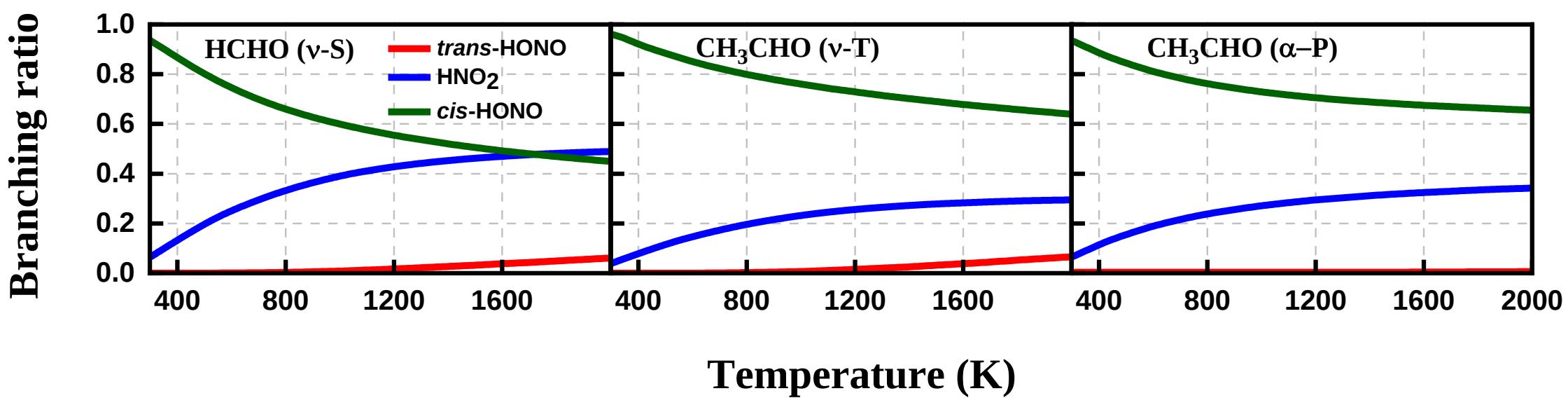

Figure 16. The comparison between the branching ratio for H-atom abstraction by $NO_2$ from aldehydes to form $HNO_2$ isomers (*trans*-HONO, $HNO_2$, *cis*-HONO), (a) HCHO (ν-S), (b) $CH_3CHO$ (ν-T), (c) $CH_3CHO$ (α-P).

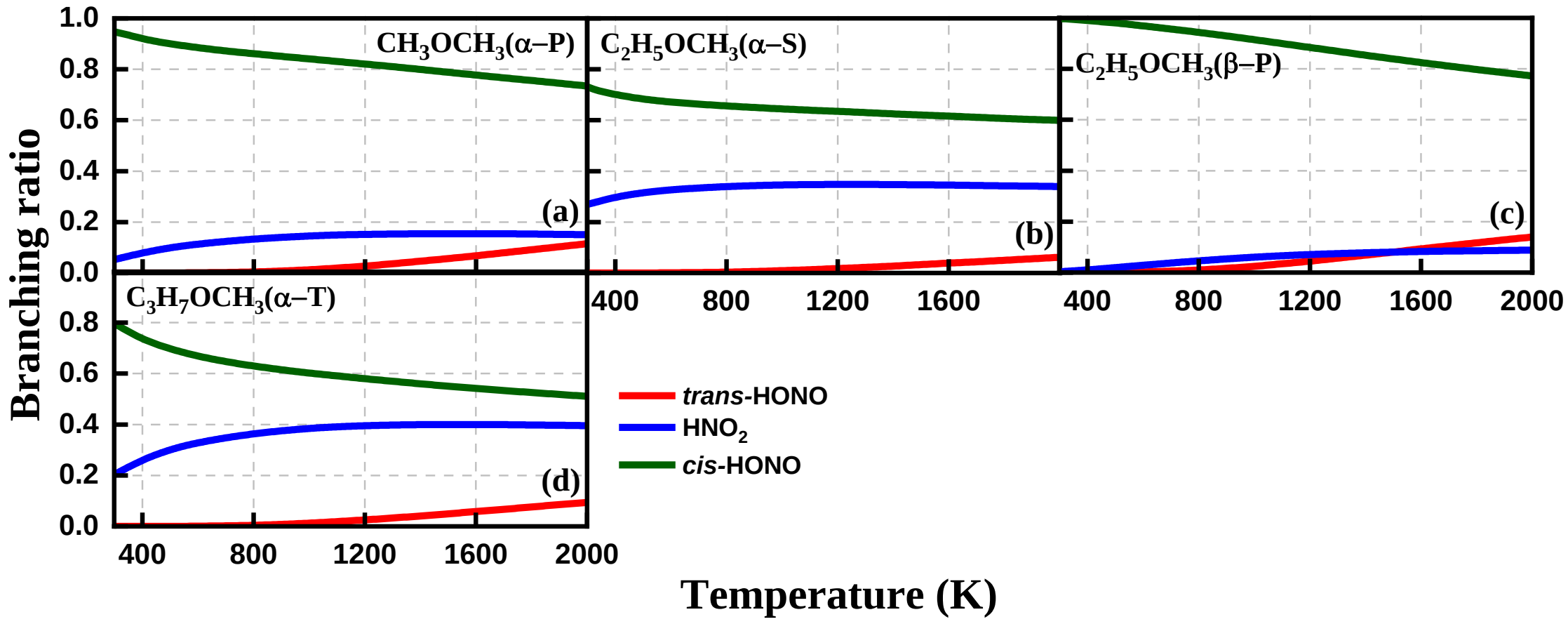

Figure 17. The comparison between the branching ratio for H-atom abstraction by $NO_2$ from ethers to form $HNO_2$ isomers (*trans*-HONO, $HNO_2$, *cis*-HONO), (a) $CH_3OCH_3$ (α-P), (b) $C_2H_5OCH_3$ (α-S), (c) $C_2H_5OCH_3$ (β-P), (d) $C_3H_7OCH_3$ (α-T).

Figures 16 and 17 further show the branching ratios for H-atom abstraction by the $NO_2$ from aldehydes and ethers. The trends of branching ratio for all reactions are similar to those observed in Fig. 15, where the *cis*-HONO channel is the most dominant. Notably, the branching ratios of *cis*-HONO and $HNO_2$ for HCHO at the ν-S site intersect at approximately 1700K (Fig. 16(a)), which coincides with the intersection of the rate coefficients for these reactions at the same temperature, Fig. S5(a). This is attributed to the enhanced influence of hindered rotor effects at higher temperature, which favors the formation of $HNO_2$.

### *3.5. Rate rules for branching ratio*

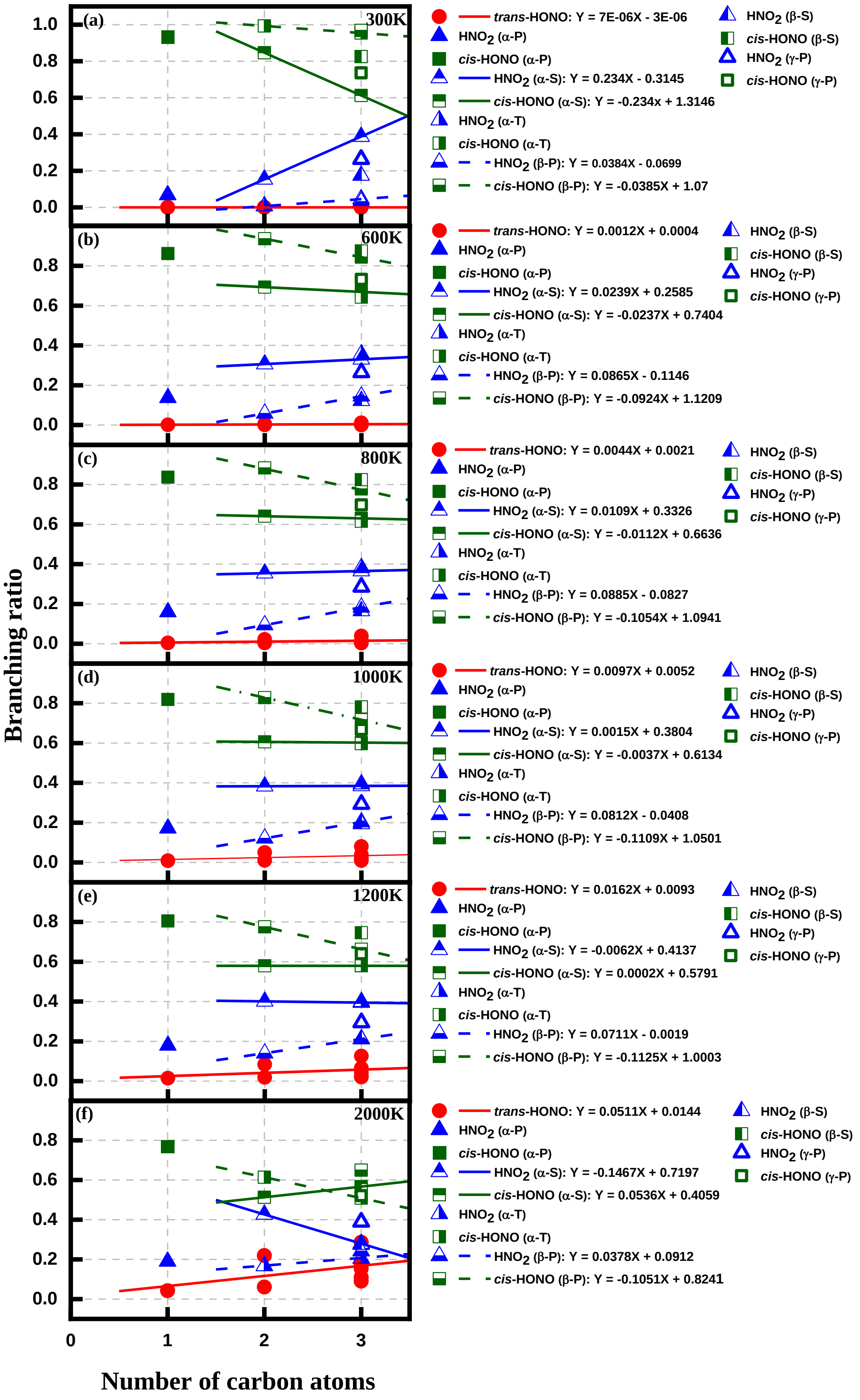

Figure 18. Branching ratio of alcohols versus carbon number at (a) 300K, (b) 600K, (c) 800K, (d) 1000K, (e) 1200K, and (f) 2000K. Lines are linear fittings, with the explicit

| fitting correlation marked on the right. |
| --- |

The consistent trends at different sites on the same molecule and at the same site on different molecules offer an opportunity to derive a consistent rule (referred to as “rate rule” in the following) to determine the branching ratios for H-atom abstraction reactions by $NO_2$ from alcohols, aldehydes and ethers at any reaction site and with any carbon number. Particularly, the branching ratios for the same type of reaction exhibit similar trends, as mentioned earlier. When comparing the branching ratios with the carbon number, a clear pattern emerges. This rate rule is also confirmed by previous study [25]. In this regard, the branching ratios at different sites of different molecules are presented as functions of carbon number at six different temperatures (i.e., 300K, 600K, 800K, 1000K, 1200K, and 2000K), which are shown in Figs. 18-19. Notably, the branching ratio of different sites at same carbon number to form same product have a distinctive difference. And the branching ratio of the same site with different carbon numbers follow a linear trend. For instance, the branching ratio of *cis*-HONO related to the carbon number can be linearized to $Y = -0.0237X + 0.7494$ and $Y = -0.0924X + 1.1209$ for the H-atom abstraction from (α-S) and (β-P) site of alcohol, respectively, Fig.18(b). And due to the unpopular of *trans*-HONO channel, the branching ratio for *trans*-HONO channel can be correlated with a single linear fit, regardless of the abstraction sites.

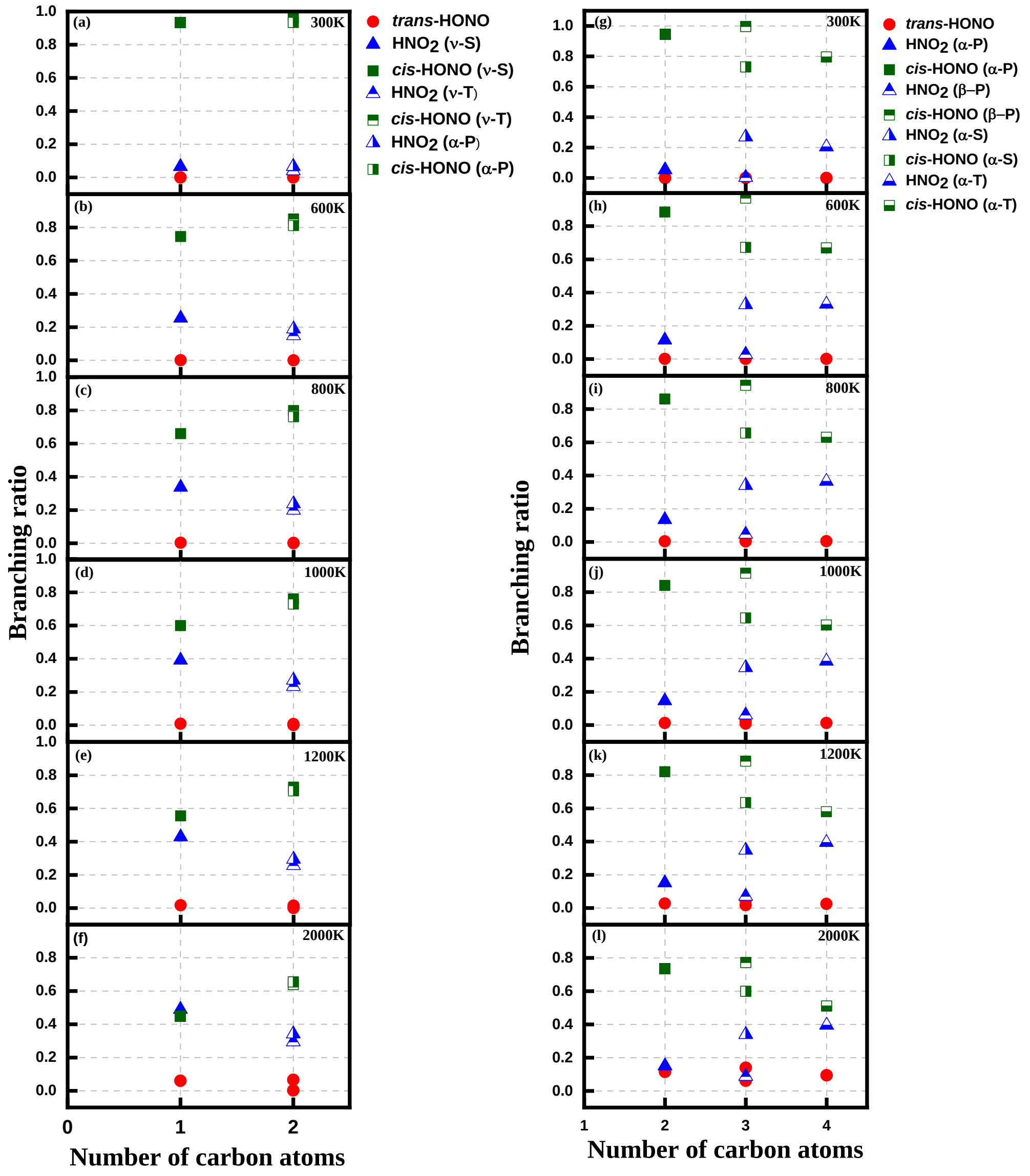

Figure 19. Branching ratio of aldehydes (left panel) and ethers (right panel) versus carbon number at 300K ((a)&(g)), 600K ((b)&(h)), 800K ((c)&(i)), 1000K ((d)&(j)), 1200K ((e)&(k)) and 2000K ((f)&(l)).

For the aldehydes, Fig.19 left panel, the trends of branching ratio represented H-atom abstraction to form *cis*-HONO is the majority, and the proportion have a rise with the increasing carbon number. Nevertheless, the branching ratio of $HNO_2$ occupies the secondary proportion and has an adverse trend compared with that of *cis*-HONO. As for ethers, Fig.19 right panel, the trends of the branching ratio for H atom abstraction to form *cis*-HONO and $HNO_2$ have an inverse direction compared to those of aldehydes. It is worth noting that the distinction of the trends of different products gradually declines following with the increasing temperature. And the channel to produce the *trans*-HONO plays an insignificant role for the whole combustion system.

*3.5 Model implementation and implications*

To demonstrate the influence caused by H-atom abstraction reactions by $NO_2$, the rate constants for each species calculated by this study are updated individually into the LLNL kinetic model [40]. Then the ignition delay time (IDT) is calculated using the kinetic model without modification and the kinetic model with newly calculated H-atom abstraction rate constants, referred to as 'original' and 'updated' hereafter. The conditions used for the NOx-doping experiments in the rapid compression machine [8] were replicated, including a temperature range of 600–1200 K, a pressure of 40 bar, an equivalence ratio of 1, and 1000 ppm $NO_2$ doping.

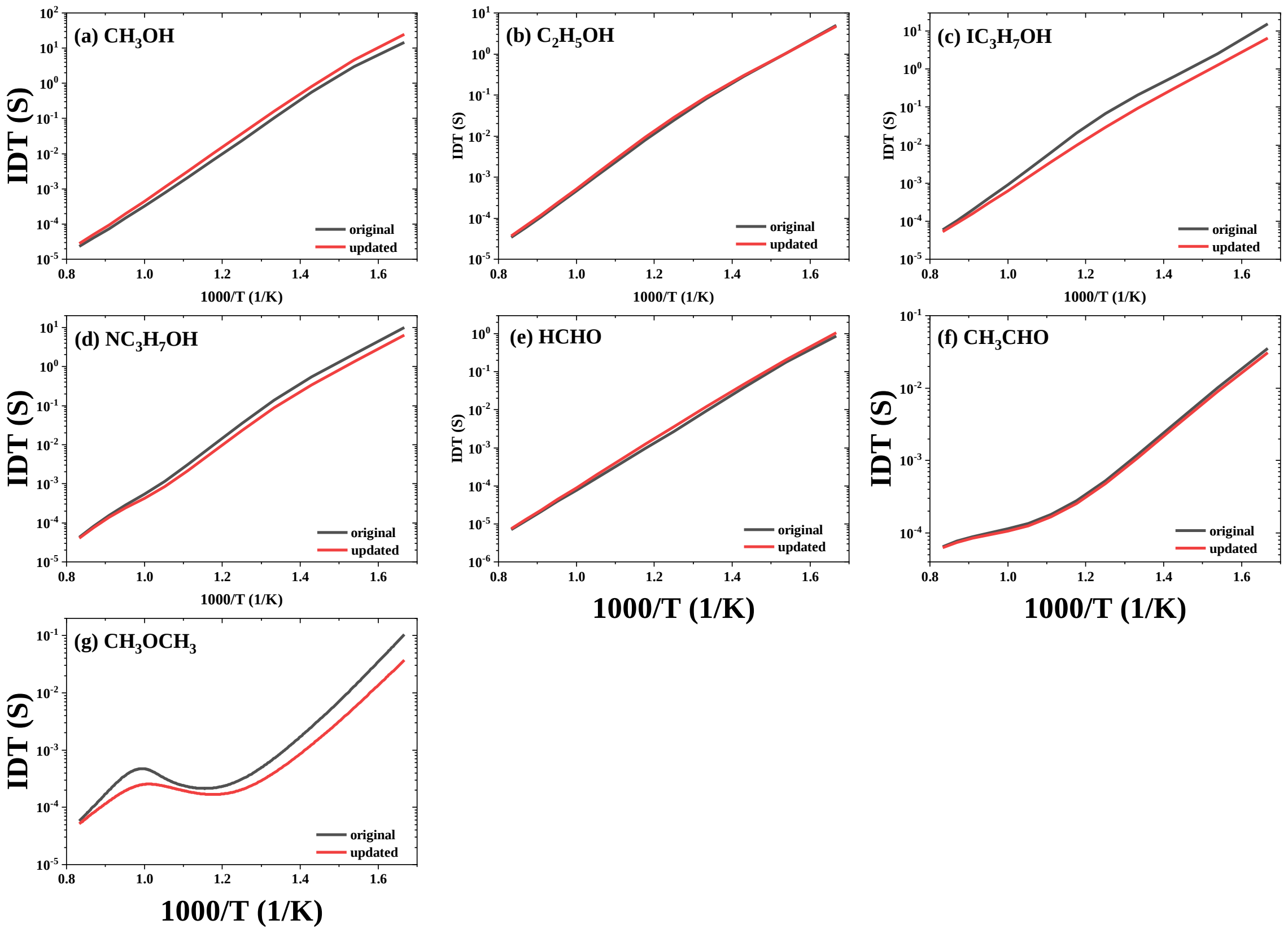

Figure 20. The IDT of investigated species calculated using the original and updated kinetic model at temperature range of 600-1200 K, pressure of 40 bar, equivalence ratio of 1, and with 1000 ppm $NO_2$ doped.

Figure 20 shows the comparison between the IDTs calculated using original and updated kinetic models for aldehydes, alcohols, and ethers. Due to the lack of a detailed reaction network for certain species, only four alcohols, two aldehydes, and one ether are investigated herein. For alcohols, the rate constants of H-atom abstraction from $CH_3OH$ in the original model are calculated by Xiao et al. [16] at CBS-Q//B3LYP/6-311++G(2d,p) level of theory. It is shown in Fig. 7 that the rate constant of $CH_3OH+NO_2 = CH_2OH_\alpha(P)+$*cis*-HONO calculated by Xiao et al. [16] is approximately one order of magnitude higher than that of this study, while the difference in rate constant for $CH_3OH+NO_2 = CH_2OH+HNO_2$ between these two studies is miner. Thus, the IDT calculated using the updated model of $CH_3OH$ is slightly longer. The rate constant of H-atom abstraction by $NO_2$ from $C_2H_5OH$ to form HONO also already exists in the original model, which is estimated by Cheng et al. [40] according to the site characteristics without distinction of HONO structure. As shown in Fig. S15, the calculated rate constant of $C_2H_5OH$+$NO_2$ = $C_2H_4OH$_β(P)+HONO by this study is two orders of magnitude lower than the estimated value, while the difference between the calculated and estimated rate constants for $C_2H_5OH$+$NO_2$ = $C_2H_4OH$_α(S)+HONO is within one order of magnitude. Due to the reason that H-atom abstraction from secondary site of $C_2H_5OH$ is the domination reaction pathway, the IDT of $C_2H_5OH$ rarely changes after incorporating the calculated rate constants to the original model. For both $IC_3H_7OH$ and $NC_3H_7OH$, the rate constants for H-atom abstraction were previously unavailable. Incorporating the calculated rate constants into the original model promotes the ignition for both species, indicated by the shortened IDT of $IC_3H_7OH$ and $NC_3H_7OH$ shown in Fig. 20 (c) and (d). For aldehyde, the rate constants of H-atom abstraction by $NO_2$ from $CH_2O$ calculated by Xu et al. [22] are adopted in the original model. As shown in Fig. 8, the rate constant of HCHO+$NO_2$ = HCO+*cis*-HONO calculated by Xu et al. [22] is about one order of magnitude higher than the value calculated by this study at low temperature range, while the rate constant of HCHO+$NO_2$ = HCO+$HNO_2$ from both studies are very similar. Therefore, substituting the rate constants for H-atom abstraction by $NO_2$ from formaldehyde (HCHO) with those calculated in this study leads to a slightly increased IDT. Unlike HCHO, the rate constants of $CH_3CHO$ are non-existent in the original model. Incorporating the rate constants for H-atom abstractions by $NO_2$ from $CH_3CHO$ to the original model slightly enhances its reactivity, resulting in a minimally shortened IDT for $CH_3CHO$. For $CH_3OCH_3$, the addition of H-atom abstraction by $NO_2$ reactions improves the reactivity of $CH_3OCH_3$ in the kinetics model, advancing the ignition of $CH_3OCH_3$, especially at low and negative temperature coefficient (NTC) end-up temperature range.

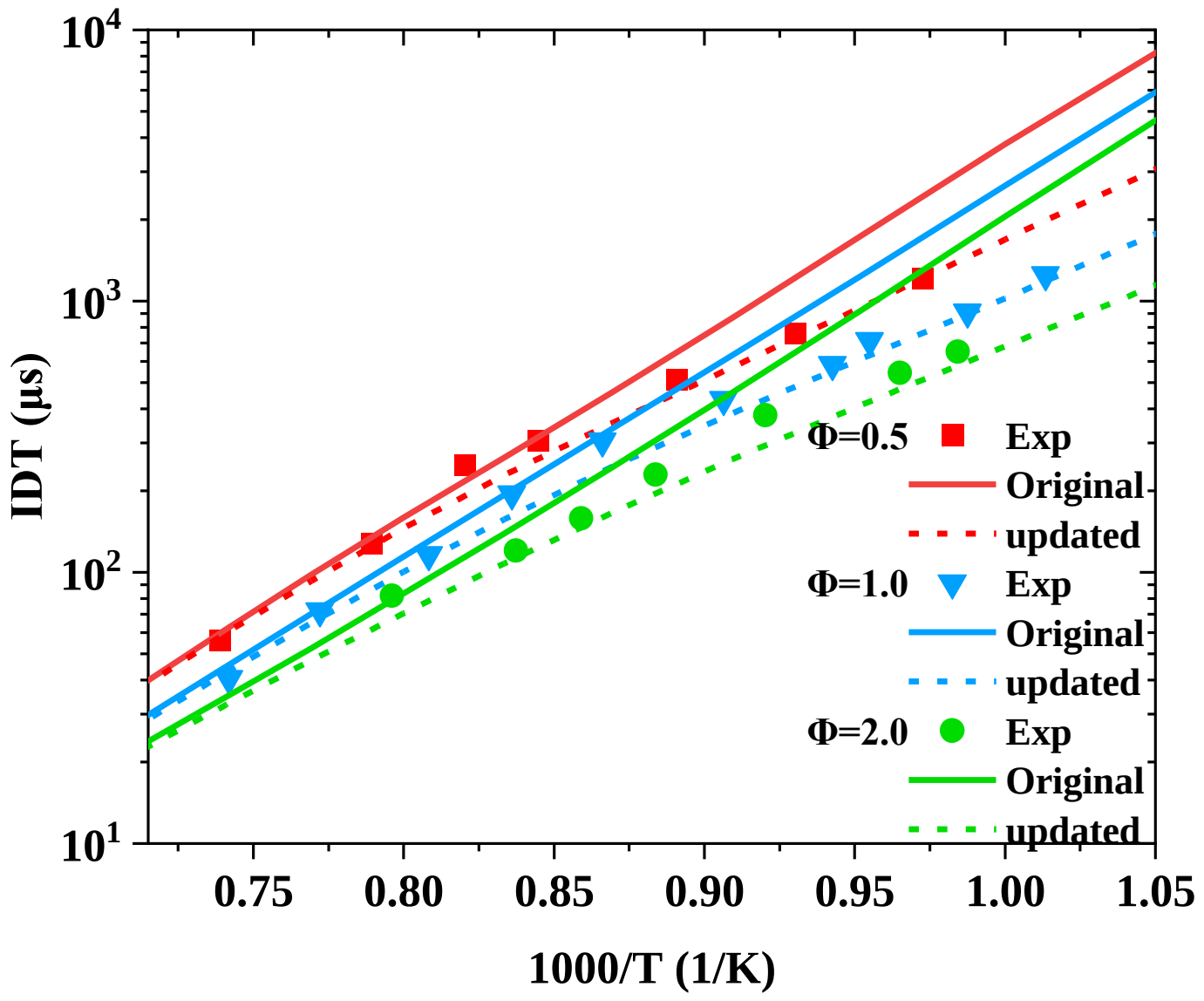

Figure 21. Comparison between the experimental IDTs for $CH_3OCH_3$ obtained from a shock tube experiment [45] and the simulated IDTs using both the updated and original kinetic models at a pressure of 10 atm, within a temperature range of 950–1400 K, and at equivalence ratios of 0.5, 1.0, and 2.0, with the $NO_2$ concentration set at 70% of the $CH_3OCH_3$ concentration.

To further demonstrate the influence caused by H-atom abstract reactions, experimental data for species that show significant changes due to the incorporation of calculated H-atom abstraction reactions are compared with the simulated results from both the original and updated models. However, experimental data for $NC_3H_7OH$ and $IC_3H_7OH$ doped with $NO_2$ are not available. Besides, experiments involving very low concentrations of $NO_2$ are excluded, as the concentration is too low to produce a noticeable effect. Therefore, only IDTs of $CH_3OCH_3$ doped with $NO_2$ obtained from shock tube experiment [45] are compared with the simulated results in Fig. 21. It was observed that, at all equivalence ratios, after incorporating the rate constants for H-atom abstraction reactions, the IDTs of $CH_3OCH_3$ were significantly shortened at intermediate to higher temperatures. This adjustment brings the simulated results into obviously closer agreement with the experimental data across the entirely investigated temperature range.

To investigate the underlying kinetics governing the model reactivity after updating the newly calculated rate constants, the sensitivity analysis is conducted for $IC_3H_7OH$, $CH_3CHO$, and $CH_3OCH_3$ under the conditions shown in Figs. 22-24, as representative for aldehyde, alcohol, and ether. The 700 K and 1100 K are selected to represent the situation at low and high temperatures respectively. The sensitivity coefficients are defined as $S_{rel} = \ln\left(\frac{\tau^{\Delta}}{\tau}\right)/\ln\left(\frac{k^{\Delta}}{k}\right)$, where $\tau^{\Delta}$ is the main IDT after multiplying the original rate constant by 2, i.e., $k^{\Delta} = 2 * k$, and $\tau$ is the original ignition delay time. The negative sensitivity coefficient indicates the promotion effect, while the positive sensitivity coefficient indicates the inhibition effect.

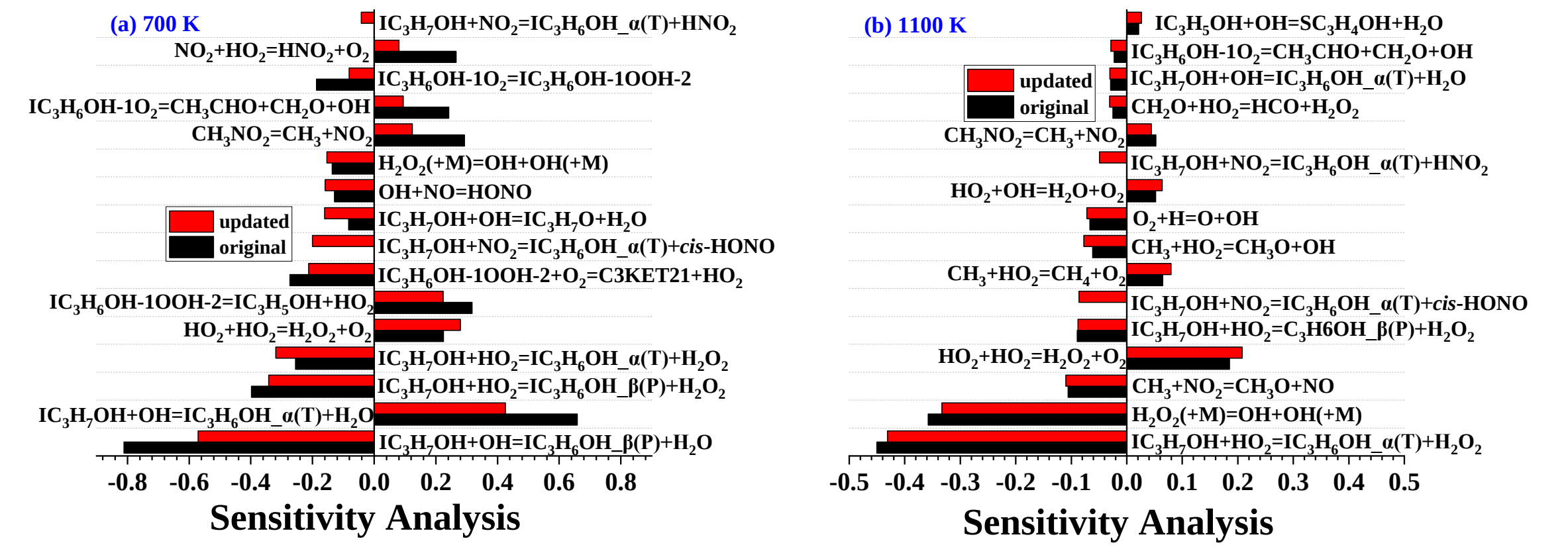

Figure 22. The sensitivity analysis on IDT for $IC_3H_7OH$ at $P_C$=40 bar, equivalence ratio of 1, and with 1000 ppm $NO_2$ doped. (a) $T_C$=700 K and (b) $T_C$=1000 K.

Figure 22 shows the calculated sensitivity coefficients on IDT for $IC_3H_7OH$. It is shown that both $IC_3H_7OH+NO_2$ = $IC_3H_6OH$_α(T)+*cis*-HONO and $IC_3H_7OH+NO_2$ = $IC_3H_6OH$_α(T)+$HNO_2$ have an obvious promotion impact on the ignition of $IC_3H_7OH$ at 700 K and 1100 K. As such, adding these H-atom abstraction reactions improves the kinetic model's reactivity and shortens the simulated IDT. The effect of most inhibition reactions decreases obviously in the updated model at 700 K, leading to a higher decrease in IDT at 700 K. Besides, the channel producing *cis*-HONO acquires the largest absolute value of the sensitivity coefficient, while the channel producing *trans*-HONO disappears in the ranking of the most impactful reactions. It is consistent with the branching ratio shown in Fig. 15 that the *cis*-HONO channel dominates at all temperatures, while the *trans*-HONO channel is at the bottom. Furthermore, it is also worth mentioning that only H-atom abstraction from the tertiary site of $IC_3H_7OH$ is shown in Fig. 22, indicating that dehydrogenation in the tertiary site is more reactive.  This is consistent with the fact that the rate constant of H-atom abstraction from the tertiary site is much higher than that from the primary site, shown in Fig. S12.

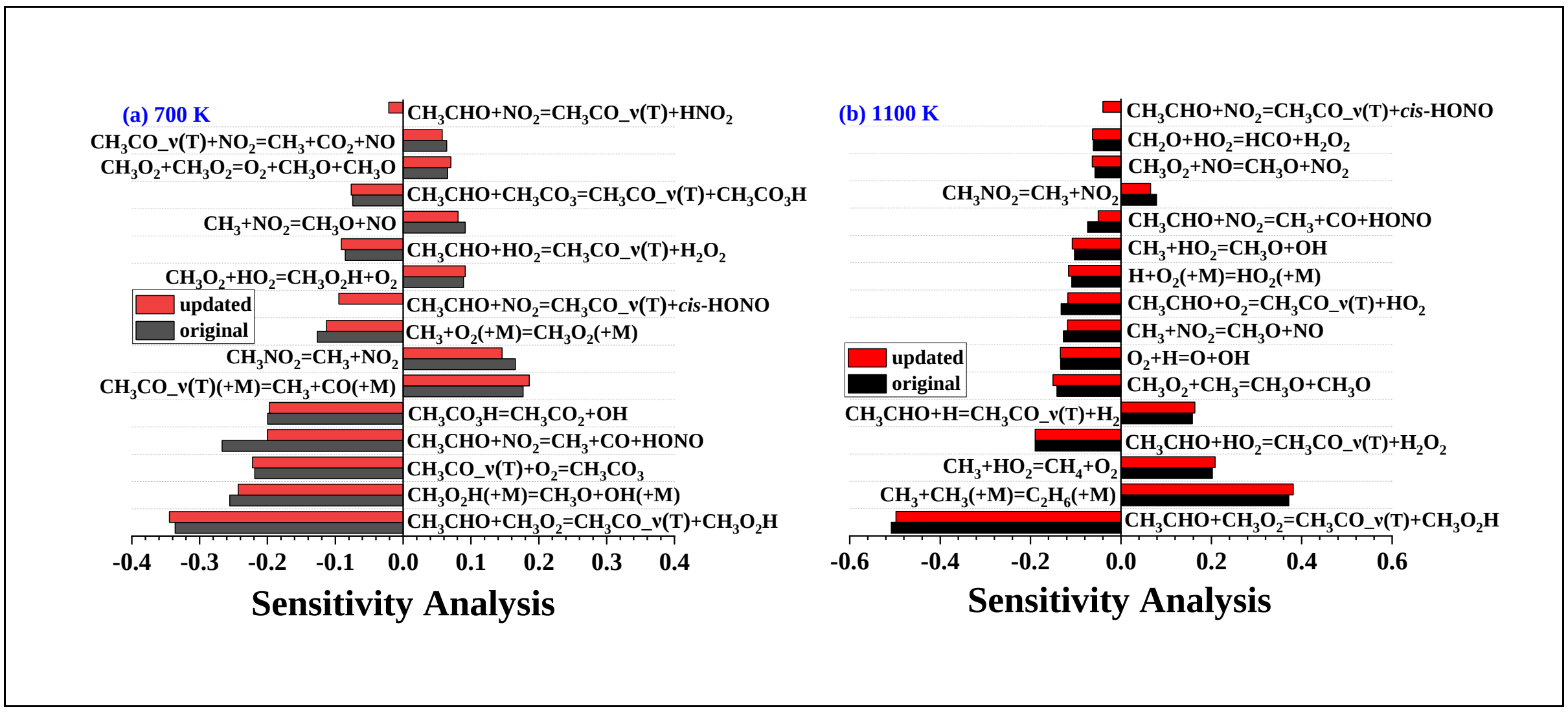

Figure 23. The sensitivity analysis on IDT for $CH_3CHO$ at $P_C$=40 bar, equivalence ratio of 1, and with 1000 ppm $NO_2$ doped. (a) $T_C$=700 K and (b) $T_C$=1000 K.

The sensitivity analysis on IDT for $CH_3CHO$ is shown in Fig. 23. Overall, the H-atom abstraction by $NO_2$ reactions show a promotional effect on $CH_3CHO$ ignition. Therefore, the addition of these reactions shortens the IDT of $CH_3CHO$. Among the calculated six reactions for H-atom abstraction from $CH_3CHO$, only $CH_3CHO$+$NO_2$ = $CH_3CO$_v(T)+*cis*-HONO and $CH_3CHO$+$NO_2$ = $CH_3CO$_v(T)+$HNO_2$ both show an obvious promotion effect at 700 K, while only $CH_3CHO$+$NO_2$ = $CH_3CO$_v(T)+*cis*-HONO is significant at 1100 K, leading to a higher IDT decrease at 700 K. This is consistent with the information given by Fig. S13 that the H-atom abstraction form ν site of $CH_3CHO$ is more reactive than from α site. Besides, it is also consistent with the branching ratio results shown in Fig. 16 that the *cis*-HONO channel is the most influential branch.

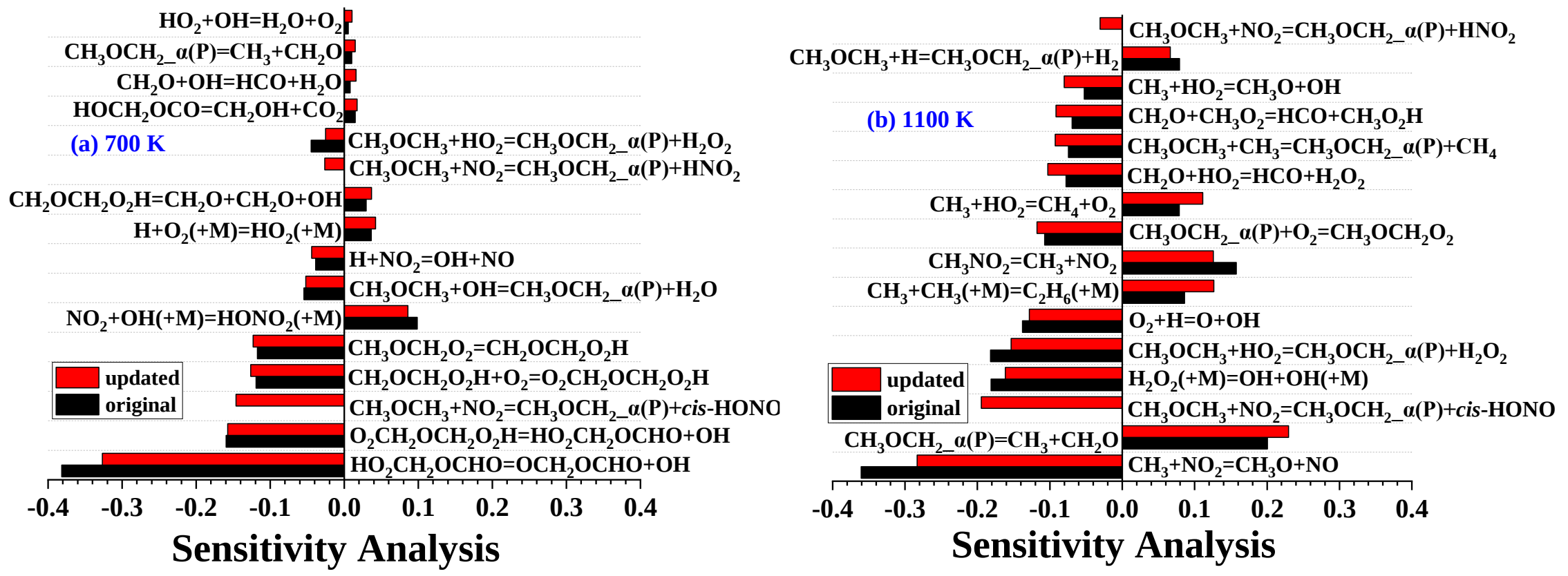

Figure 24. The sensitivity analysis on IDT for $CH_3OCH_3$ at $P_C$=40 bar, equivalence ratio of 1, and with 1000 ppm $NO_2$ doped. (a) $T_C$=700 K and (b) $T_C$=1000 K.

Figure 24 shows the sensitivity analysis on IDT for $CH_3OCH_3$. It is found that the $CH_3OCH_3$+$NO_2$ = $CH_3OCH_2$_α(P)+*cis*-HONO and $CH_3OCH_3$+$NO_2$ = $CH_3OCH_2$_α(P)+$HNO_2$ promote the ignition of $CH_3OCH_3$. Moreover, $CH_3OCH_3$+$NO_2$ = $CH_3OCH_2$_α(P)+*cis*-HONO is the third-most promotion reaction at both temperatures. Adding these H-atom abstraction reactions significantly shortens the IDT of $CH_3OCH_3$ across both low and NTC end-up temperature ranges. Similar to $CH_3CHO$ and $IC_3H_7OH$, the *cis*-HONO channel plays the most important role. This is also supported by the branching ratios shown in Fig. 17, where *cis*-HONO is the most dominant pathway.

Flux analyses are further conducted to demonstrate the underlying reasons under the same condition and for the same species as the sensitivity analysis, specifically at 0.1% fuel consumption. The results are summarized in Figs. 25-27. The percentages represent the ratio of the consumption rate for that pathway to the total consumption rate.

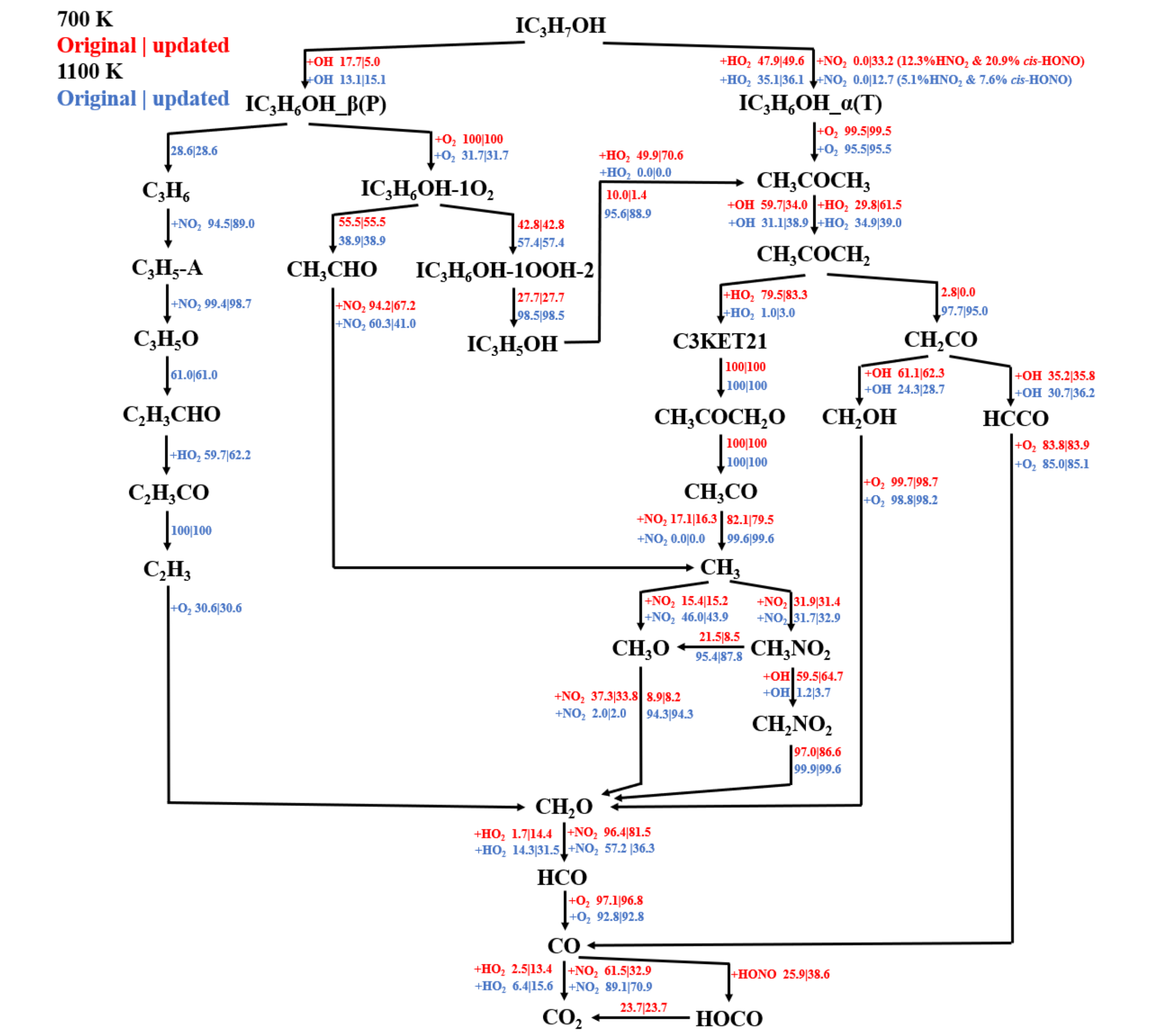

Figure 25. The flux analysis results for $IC_3H_7OH$ ignition at $P_C$ = 40 bar, $T_C$ = 750 and 1100 K, equivalence ratio of 1, and with 1000 ppm $NO_2$ doped. Fluxes are computed at 0.1% $IC_3H_7OH$ consumption.

Figure 25 shows the flux analysis results for $IC_3H_7OH$. It is found that the $NO_2$ participates in the dehydrogenation of $IC_3H_7OH$ obviously in the updated model, with 33.7% and 12.7% at 700 K and 1100 K. It is worth mentioning that it only happens at the tertiary site of $IC_3H_7OH$, being consistent with the sensitivity analysis. As significant promotion reactions, they promote the initiation of fuel oxidation and advance the ignition. Additionally, the higher percentage of $IC_3H_7OH$ consumed by $NO_2$ at 700 K compared to 1100 K aligns with the observation that IDT is more strongly promoted at 700 K than at 1100 K. Besides, the *cis*-HONO and $HNO_2$ are the main products of H-atom abstraction by $NO_2$, with *cis*-HONO being a dominant product. This confirms that *cis*-HONO is the most significant product, as also indicated by the branching ratio results and sensitivity analysis for $IC_3H_7OH$.

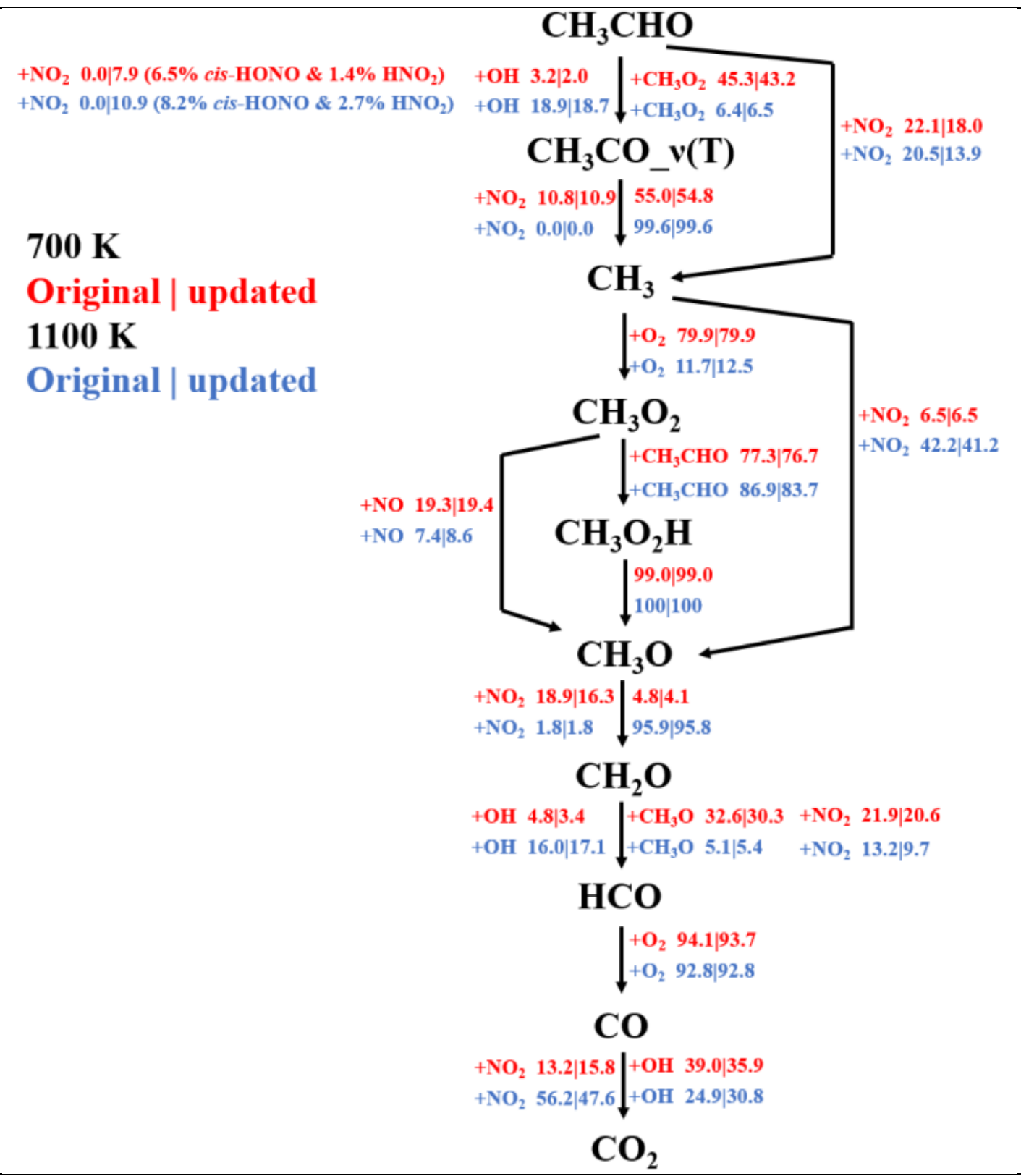

Figure 26. The flux analysis results for $CH_3CHO$ ignition at $P_C$ = 40 bar, $T_C$ = 750 and 1100 K, equivalence ratio of 1, and with 1000 ppm $NO_2$ doped. Fluxes are computed at 0.1% $CH_3CHO$ consumption.

As shown in Fig. 26, the inclusion of H-atom abstraction by $NO_2$ reactions allows $NO_2$ to participate in the dehydrogenation of fuel molecules, with 7.9% and 10.9% of $CH_3CHO$ being consumed by $NO_2$ at 700 K and 1100 K, respectively. Conversely, the original pathway for $NO_2$ reacting with $CH_3CHO$, i.e., $CH_3CHO+NO_2 = CH_3+CO+HONO$, shows a reduced impact, with decreases of 4.1% and 6.6% at 700 K and 1100 K, respectively. Both the H-atom abstraction by $NO_2$ and the original pathway are classified as promoting reactions, as indicated by sensitivity analysis. The former enhances its promoting effect, while the latter diminishes it. Consequently, the ignition of $CH_3CHO$ is slightly advanced. Furthermore, most of the products from the reaction of H-atom abstraction by $NO_2$ are *cis*-HONO, with 6.5% producing *cis*-HONO and 1.4% producing $HNO_2$ at 700 K, and 8.2% producing *cis*-HONO and 2.7% producing $HNO_2$ at 1100 K. This is supported by the branching ratio results and sensitivity analysis, which indicate that the *cis*-HONO pathway is the dominant channel and followed by $HNO_2$ pathway.

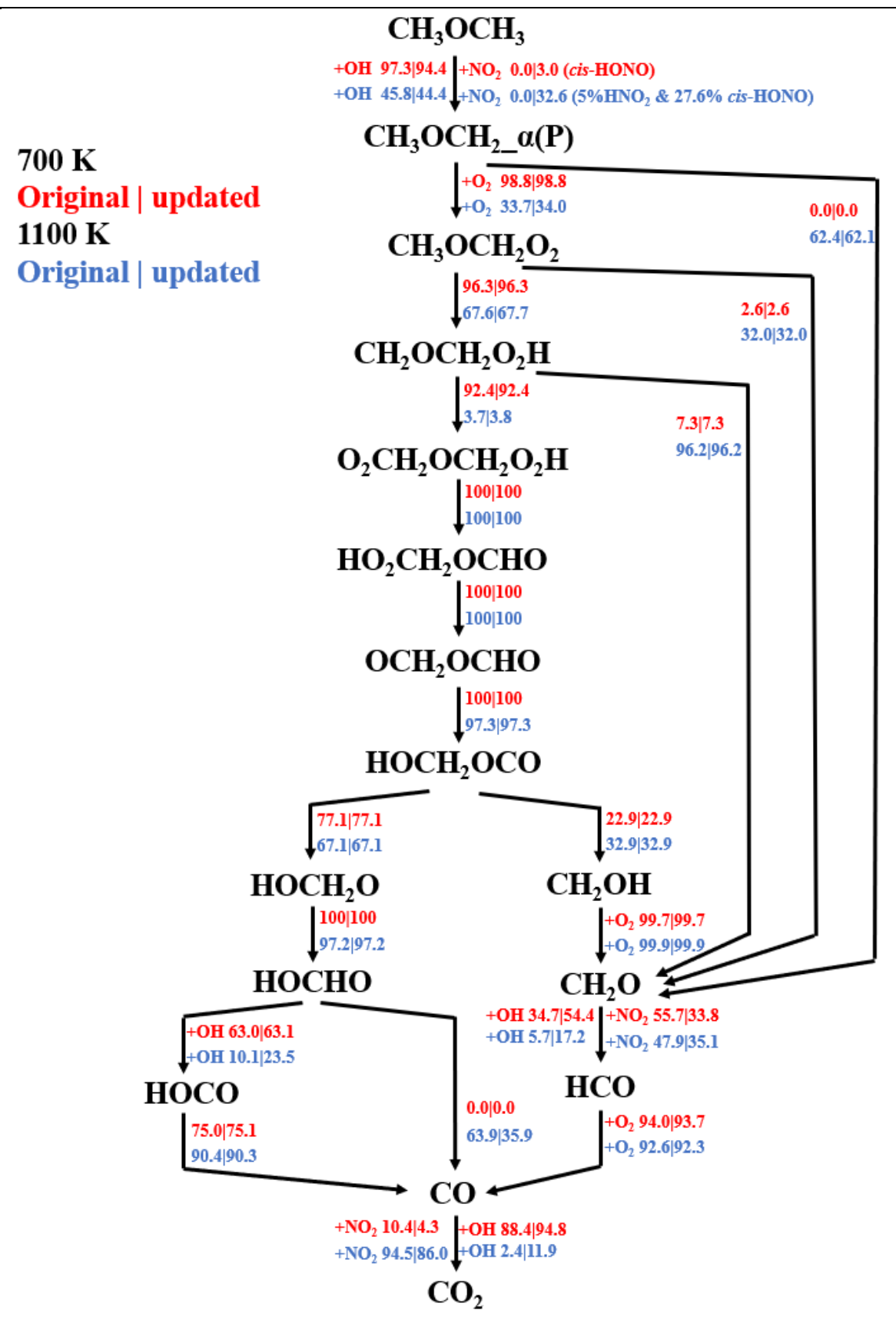

Figure 27. The flux analysis results for $CH_3OCH_3$ ignition at $P_C$ = 40 bar, $T_C$ = 750 and 1100 K, equivalence ratio of 1, and with 1000 ppm $NO_2$ doped. Fluxes are computed at 0.1% $CH_3OCH_3$ consumption.

The flux analysis results for $CH_3OCH_3$ are shown in Fig. 27. It is found that 3% of $CH_3OCH_3$ is consumed by $NO_2$ with *cis*-HONO produced at 700 K, indicating that $NO_2$ abstracting H-atom starting to take effect on $CH_3OCH_3$. Although the sensitivity analysis shows that the $HNO_2$ channel also exhibits a notable promotion effect, this is not supported by the flux analysis. As for 1100 K, 32.6% of $CH_3OCH_3$ is consumed by $NO_2$. This high consumption percentage supports the observation that IDT of $CH_3OCH_3$ is still significantly promoted by H-atom abstraction by $NO_2$. Furthermore, the *cis*-HONO pathway is still observed as the dominant pathway in the flux analysis of $CH_3OCH_3$.

## 4. Conclusion

This work presents a systematic investigation of H-atom abstractions by $NO_2$ from different sites of $C_1$-$C_4$ alcohols, aldehydes, and ethers, leading to the formation of three $HNO_2$ isomers (*trans*-HONO, $HNO_2$, and *cis*-HONO). The geometry optimizations and vibrational frequency calculations of all involved species are conducted at M06-2X/6-311++G(d,p) level of the theory, while single-point energies (SPEs) are calculated using the DLPNO-CCSD(T)/cc-pVDZ method. Energy barriers for 45 reactions are determined and analyzed in conjunction with the corresponding bond dissociation energies (BDEs). Temperature-dependent rate coefficients for these 45 reactions are proposed using the Master Equation System Solver (MESS) [36] over a temperature range of 298.15–2000 K, based on conventional transition state theory with unsymmetric Eckart tunneling corrections. These updated rate constants are incorporated into a recently enhanced chemical kinetic model, and their effects on model performance are evaluated through comprehensive kinetic modeling. Sensitivity and flux analyses are performed to further investigate the chemical kinetics that influence changes in model performance. The primary conclusions from this study are:

- The energy barriers at different sites of alcohols and ethers follow a similar ranking. In alcohols, the barriers are ranked as β(P) > γ(P) > β(S) > α(P) > α(S) > α(T), while in ethers, the order is β(P) > α(P) > α(S) > α(T). The C-O functional groups significantly lower the energy barriers for abstracting the adjacent hydrogen atoms at the α carbon site. For aldehydes, H-atom abstractions at the C=O sites exhibit markedly lower energy barriers than at all other sites. This trend contrasts with the energy barriers for H-atom abstractions at C=C sites, as reported in the author's previous research [25]. Across all reaction sites studied, the energy barriers for producing *trans*-HONO are consistently higher than those forming $HNO_2$ and *cis*-HONO.
- The branching ratios of the pathways forming $HNO_2$, *cis*-HONO and *trans*-HONO vary between different species and between different carbon sites on the same molecule, with the *cis*-HONO-producing pathway being the most dominant for most species. Nevertheless, further analysis indicates that the branching ratios exhibit clear trends on carbon numbers for different channels. Different rate rules, as demonstrated in a previous study [25], have been proposed for various sites. These rules can be used to analogize rate coefficients for this type of reactions to heavier hydrocarbons (e.g., >$C_4$) at various reaction sites.
- The H-atom abstraction by $NO_2$ accelerates the initial oxidation for fuel molecules and enhances the autoignition reactivity of the investigated species within the kinetic model. This promotion effect is particularly pronounced at low temperatures for aldehydes and alcohols. The highest promotion impact on ignition reactivity is observed for $CH_3OCH_3$, where H-atom abstraction by $NO_2$ not only advances the IDT within the low-temperature regime but also at the NTC temperature range. Incorporating these reactions into the kinetic model leads to greatly improved agreement between modeling results and experimental measurements.

- Both the sensitivity and flux analyses highlight the critical role of the H-atom abstraction by $NO_2$ from fuel molecules and critical intermediates (e.g., HCHO, HCO) in determine the branching ratio of fuel consumption pathways, with the *cis*-HONO pathway being the dominant one, followed by the $HNO_2$ pathway, then the *trans*-HONO pathway.

## Acknowledgments

The work described in this paper is supported by the Research Grants Council of the Hong Kong Special Administrative Region, China under PolyU P0046985 for ECS project funded in 2023/24 Exercise, the Otto Poon Charitable Foundation under P0050998, the National Natural Science Foundation of China under 52406158, the Chief Executive's Policy Unit of HKSAR under the Public Policy Research Funding Scheme (2024.A6.252.24B), and the Natural Science Foundation of Guangdong Province under 2023A1515010976 and 2024A1515011486.

## Declaration of Competing Interests

The authors declare no competing interests.